\documentclass[11pt]{article}
\usepackage{fullpage}
\usepackage[numbers]{natbib}
\usepackage[margin=1in]{geometry}
\usepackage{times}
\usepackage[compact]{titlesec}

\usepackage{booktabs} % For formal tables

\usepackage{hyperref}       % hyperlinks
\usepackage{url}            % simple URL typesetting
\usepackage{amsfonts}       % blackboard math symbols
\usepackage{nicefrac}       % compact symbols for 1/2, etc.
\usepackage{microtype}      % microtypography
\usepackage{xcolor}% colors
\usepackage{amsmath,amsthm,amssymb}
\usepackage{latexsym}
\usepackage{epic}
\usepackage{epsfig}
\usepackage{verbatim}
\usepackage[justification=centering]{caption}
\usepackage{enumitem}
\usepackage{color}
\usepackage{bbm}
\usepackage[normalem]{ulem}
\usepackage{thmtools,thm-restate}
\usepackage{float}
\usepackage{tikz}
\usepackage{multirow}
\usepackage[normalem]{ulem}
\usepackage{mathtools}
\usepackage[ruled, vlined, linesnumbered]{algorithm2e}
\SetAlFnt{\small}
\SetAlCapFnt{\small}
\SetAlCapNameFnt{\small}
\SetAlCapHSkip{0pt}
\IncMargin{-\parindent}

\newtheorem{theorem}{Theorem}

\newtheorem{corollary}{Corollary}
\newtheorem{lemma}{Lemma}

\newtheorem{proposition}{Proposition}

\newtheorem{definition}{Definition}

\newtheorem{example}{Example}
\newtheoremstyle{nonindented}{1ex}{1ex}{}{}{\bfseries}{.}{.5em}{}
\newtheoremstyle{indented}{1ex}{1ex}{\itshape\addtolength{\leftskip}{0.6cm}\addtolength{\rightskip}{0.6cm}}{}{\bfseries}{.}{.5em}{}
\theoremstyle{nonindented}
\theoremstyle{indented}
\theoremstyle{plain}

\usepackage{graphicx}
\usepackage[ruled, vlined, linesnumbered]{algorithm2e}
\usepackage[numbers]{natbib}
\usepackage{amsmath,amssymb}
\usepackage{color}
\usepackage{xfrac}
\usepackage{hyperref}

\usepackage{bm}

\newcommand{\bvec}[1]{\boldsymbol{ #1 }}

\newcommand{\union}{\cup}

\renewcommand{\hat}{\widehat}
\renewcommand{\tilde}{\widetilde}
\renewcommand{\bar}{\overline}

\DeclareMathOperator{\poly}{poly}

\def\min{\qopname\relax n{min}}
\def\max{\qopname\relax n{max}}
\def\argmin{\qopname\relax n{argmin}}
\def\argmax{\qopname\relax n{argmax}}

\def\Pr{\qopname\relax n{\mathbf{Pr}}}
\def\Ex{\qopname\relax n{\mathbf{E}}}

\newcommand{\RR}{\mathbb{R}}

\newcommand{\cX}{\mathcal{X}}

\newcommand{\eat}[1]{}

\newcommand{\maxi}[1]{\mbox{maximize} & {#1 } & \\}

\newcommand{\st}{\mbox{subject to} }

\newcommand{\con}[1]{&#1 & \\}
\newcommand{\qcon}[2]{&#1, & \mbox{for } #2.  \\}
\newenvironment{lp}{\begin{equation}  \begin{array}{lll}}{\end{array}\end{equation} }
\newenvironment{lp*}{\begin{equation*}  \begin{array}{lll}}{\end{array}\end{equation*}}

\newcommand{\norm}[1]{||#1||}
\newcommand{\BR}{\operatorname{BR}}

\definecolor{darkgreen}{rgb}{0.09, 0.45, 0.27}

\newcommand{\rev}[1]{{#1}}

\begin{document}

% \title{Robust Stackelberg Equilibria\thanks{An earlier version of this work was presented at the ACM Conference on Economics and Computation (EC'23) and a 1-page abstract appeared in the EC'23 conference proceedings.}
% }

% Title. Note the optional short title for running heads. In the interest of anonymization, please do not include any acknowledgements.
\title{Robust Stackelberg Equilibria\footnote{This work is supported in part by an Army Research
Office Award  W911NF-23-1-0030, NSF award CCF-2303372 and Office of Naval Research Award N00014-23-1-2802.}}

% Anonymized submission.
 % \author{Submission 83}

% \author{
% % Submission 861
% Jiarui Gan \\ University of Oxford \\ \small jiarui.gan@cs.ox.ac.uk
% \and
% Minbiao Han \\ University of Chicago \\ \small minbiaohan@uchicago.edu
% \and
% Jibang Wu \\ University of Chicago \\ \small wujibang@uchicago.edu
% \and
% Haifeng Xu\\ University of Chicago \\ \small haifengxu@uchicago.edu
% }

\author{Jiarui Gan\thanks{Department of Computer Science, University of Oxford, Oxford, UK; Email: \texttt{jiarui.gan@cs.ox.ac.uk}.} 
 \and   Minbiao Han\thanks{Department of Computer Science, University of Chicago, Chicago, IL 60637, USA; Email: \texttt{\{minbiaohan, wujibang, haifengxu\}@uchicago.edu}.}  \and  Jibang Wu\footnotemark[3] \and  Haifeng Xu\footnotemark[3]}

\begin{titlepage}

   \maketitle
   
   % Abstract. Note that this must come before \maketitle.
   \begin{abstract}
This paper provides a systematic study of the {\em robust Stackelberg equilibrium} (RSE), which  naturally extends the widely adopted solution concept of the strong Stackelberg equilibrium (SSE). The RSE accounts for \emph{any} possible up-to-$\delta$ suboptimal follower responses in Stackelberg games and is adopted to improve the robustness of the leader's strategy through worst-case analysis. While a few variants of robust Stackelberg equilibrium have been considered in the previous literature, the RSE solution concept we consider is importantly different --- in some sense, it relaxes previously studied robust Stackelberg strategies and is applicable to much broader sources of uncertainties.  
We provide a thorough investigation of several fundamental properties of RSE, including its utility guarantees, algorithmics, and learnability.  
We first show that the RSE always exists and is thus well-defined. Then we characterize how the leader's utility in RSE changes with the robustness level considered.   
On the algorithmic side, we show that, in sharp contrast to the tractability of computing an SSE, it is NP-hard to obtain a fully polynomial approximation scheme (FPTAS) for any constant robustness level.
Nevertheless, we develop a quasi-polynomial approximation scheme (QPTAS) for RSE.  
Finally, we examine the learnability of the RSE in a natural learning scenario, where both players' utilities are not known in advance, and provide almost tight sample complexity results on learning the RSE. As a corollary of this result, we also obtain an algorithm for learning SSE, which strictly improves a key result of \citet{bai2021sample} in terms of both utility guarantee and computational efficiency. 
   \end{abstract}
   
   \end{titlepage}

\section{Introduction}
The Stackelberg game stands as a cornerstone in the realm of hierarchical decision-making processes, serving as a canonical model that underpins fundamental economic models.
The game in its normal form (see Section~\ref{sec:prelim} for details) involves two players, a leader and a follower, with utility function $u_l, u_f: [m]\times [n] \to \RR$. 
The leader moves first by committing to a (possibly randomized) strategy $\bvec{x} \in \Delta^{m}$, the follower then responds with an action $j \in [n]$. Denote $u(\bvec{x}, j) = \Ex_{i\sim \bvec{x}}[ u(i,j)] $ as the player's expected payoff under utility function $u$ and strategy profile $(\bvec{x}, j)$.
A conventional design of the optimal leader strategy can be formulated as a solution to the following (optimistic) bi-level optimization problem,
\begin{equation}\label{eq:SSE}
         \max_{\bvec{x} \in \Delta^{m}} \max_{j \in [n]} u_l(\bvec{x}, j) 
       \quad \text{ {s.t.} } \quad 
        u_f(\bvec{x}, j) \geq \max_{j'\in [n]} u_f(\bvec{x}, j'),
\end{equation}
which solves for the leader payoff-maximizing strategy given that the follower would respond optimally and break ties optimistically.
This solution concept is known as the strong Stackelberg equilibrium (SSE). In particular, as long as the game instance is non-degenerated, the ``strong'' assumption of the follower's optimistic tie-breaking behavior is without loss of generality, and SSE forms the subgame perfect Nash equilibria of the underlying extensive-form game~\cite{von2010leadership}. 
The SSE is a fundamental game-theoretical concept, as it reduces to the minimax strategy in zero-sum games and is viewed as another natural extension of minimax strategy to general sum games, analogous to the Nash equilibrium~\cite{conitzer2016stackelberg}.  

However, the SSE also exhibits a critical limitation that if the follower responds (even slightly) {\em suboptimally} to the leader, the quality of the leader's SSE strategy may deteriorate substantially. These scenarios commonly occur when there is uncertainty regarding the follower's utilities  \citep{letchford2009learning,kiekintveld2013security,kroer2018robust}, or when the follower is boundedly rational \citep{yang2012computing}, or when the follower has imperfect perception of the leader's strategies \citep{an2012security,muthukumar2019robust}, or even a mixture of some of these factors \citep{pita2010robust}. In this paper, we take an agnostic stance on the cause of suboptimal follower responses and adopt a worst-case analysis for the robust design of the optimal leader strategy. 
This leads to the following (pessimistic) bi-level optimization problem,    
\begin{equation}\label{eq:RSE}
         \max_{\bvec{x} \in \Delta^{m}} \min_{j \in [n]} u_l(\bvec{x}, j) 
       \quad \text{ {s.t.} } \quad 
       u_f(\bvec{x}, j) > \max_{j'\in [n]} u_f(\bvec{x}, j') - \delta,
\end{equation}
which solves for the leader payoff-maximizing strategy given that the follower's response is the worst to the leader's utility among all actions whose utilities are up to a small difference $\delta$ from the optimal response. 
We refer to this solution concept as the $\delta$-robust Stackelberg equilibrium ($\delta$-RSE). The parameter $\delta$ reflects the follower's level of suboptimality and can be chosen according to the designer's knowledge of the game, e.g., the confidence bound on utility estimation or the degree of irrationality. This solution concept naturally extends the SSE in the sense that, as $\delta \to 0$, the leader utility of $\delta$-RSE approaches (continuously under certain conditions) \rev{towards} the leader utility of SSE. This property notably allows $\delta$-RSE to serve as the pessimistic design choice in the process of learning SSE, where $\delta$ is determined by the confidence bound in parameter estimation. We defer the detailed discussion to Section~\ref{sec:RSE-Property} and \ref{sec:learnability}.

Despite its simplicity and power to capture a wide range of suboptimal behaviors, there appears to be a lack of thorough analysis in the literature for such a natural solution concept of robust Stackelberg equilibrium.
For simultaneous move games, somewhat similar concepts of robust Nash equilibrium have been proposed and thoroughly analyzed  \citep{tijs1981nash,aghassi2006robust}. However, the analysis in simultaneous-move games is quite different from that of Stackelberg games. 
For Stackelberg games, most relevant to ours is perhaps an applied study by \citet{pita2010robust} who proposed a mixed integer linear program for computing a subtle {variant} of our RSE solution in order to find robust leader strategies. However, the running time of their program is exponential in the worst case; moreover, as we will elaborate in Section~\ref{sec:RSE-Property}, the RSE variant they considered may not always exist. Many fundamental questions still remain: How to define the equilibrium concept so that it always exists? What is the computational complexity of finding the RSE? Are there provably sub-exponential algorithms for computing the RSE? What are the connections between RSE and other equilibrium concepts? These are the questions we aim to answer in this paper.

\subsection{Our Contributions}
This paper is dedicated to a principled study of $\delta$-RSE from its analytic properties, to its computational and statistical complexity.
We start by formalizing the notion of $\delta$-RSE, where $\delta > 0$ is an upper bound on the follower's utility loss due to suboptimal behavior.
We prove that under a properly chosen boundary condition, a $\delta$-RSE exists in every Stackelberg game for any $\delta > 0$. Consequently, the leader's utility in an $\delta$-RSE can be viewed as a well-defined function $u_{\textup{RSE}}(\delta)$ of $\delta > 0$.
By analyzing the function $u_{\textup{RSE}}(\delta)$, we show several analytical properties of the $\delta$-RSE and compare it with other solution concepts, including the classic SSE and the maximin equilibrium.
In particular, under a minor non-degeneracy assumption, the function $u_{\textup{RSE}}(\delta)$ is proved to exhibit Lipschitz continuity within a regime of small $\delta$, and approaches the leader utility under SSE as $\delta \to 0$. Such   continuity   is a ``surprisingly'' nice property, as an agent's equilibrium utility typically is \emph{not} continuous in the other agents' parameters in discrete strategic games. For instance,  in SSE, the leader's utility may drop significantly if the follower's utility function changes even slightly or the follower's response is slightly sub-optimal because these may cause the follower to switch to a response that is dramatically worse to the leader. Interestingly, our added layer of worst-case analysis somehow smoothed the follower response and installed the continuity property.  This continuity property has multiple interesting implications. First, it implies that regardless of what causes the follower's up-to-$\delta$ suboptimal behavior,\footnote{It is easy to see that any $\delta$ perception error on follower's payoffs \citep{letchford2009learning,kiekintveld2013security} or leader's mixed strategy \citep{an2012security,muthukumar2019robust} will lead to $O(\delta)$ suboptimal follower responses when payoffs are bounded.} the leader's utility will always be $O(\delta)$ off from the SSE utility for small $\delta$.  Second, this property turns out to be very useful for learning both the RSE and SSE, which we will further elaborate on below.

Next, we investigate the complexity of computing a $\delta$-RSE. In sharp contrast to the tractability of computing an SSE, we show that for any $\delta>0$, it is NP-hard to even approximate (the leader's strategy in) a $\delta$-RSE;  this inapproximability result rules out the possibility of a fully polynomial time approximation scheme (FPTAS), assuming P $\neq$ NP. Our proof employs a highly nontrivial reduction from the {\sc eXact 3-set Cover (X3C)}  problem. The reduction is combinatorial in nature despite computing (continuous) mixed strategies, and a key technical challenge in the proof is to relate the continuous game strategy space to the combinatorial solution space of {\sc X3C}.  
On the positive side, we present a quasi-polynomial approximation
scheme (QPTAS) to compute an approximate RSE.  
Our proof employs the \emph{probabilistic method} \citep{alon2016probabilistic} to prove the existence of an approximate $\delta$-RSE, which is similar to \citep{lipton2003playing} for proving the existence of an approximate Nash equilibrium with a simple format termed \emph{uniform strategy} (which can be enumerated in quasi-polynomial time). However, our algorithm for identifying the approximate $\delta$-RSE is significantly different --- in fact, the approximate $\delta$-RSE is not even a uniform strategy, but instead is some strategy nearby. This is due to the nature of bi-level optimization in Stackelberg games. While a uniform leader strategy $\bar{ \bvec{x} }$  and any nearby strategy $\bvec{x}$ will lead to similar leader utilities, they will lead to different follower responses, which in turn affects what leader utility is induced in the equilibrium. This challenge forces us to efficiently search the nearby region of a uniform strategy $\bar{ \bvec{x} }$  for a leader strategy $\bvec{x}$ that induces the most favorable follower response. This challenge brought by follower responses is not present in computing an approximate Nash equilibrium. We remark that it is an intriguing yet highly non-trivial open question to close the gap between the above hardness result and QPTAS.\footnote{Familiar readers may recall that there was a similar gap in the complexity landscape of computing a Nash equilibrium, which was an open problem for about 10 years. Specifically, around 2005,  \citet{lipton2003playing} developed a QPTAS for finding an $\epsilon$-Nash whereas  \citet{daskalakis2009complexity,chen2009settling} ruled out FPTAS for two-player Nash (assuming PPAD$\not \subseteq$P). The gap between these two results was open for about 10 years until \citet{rubinstein2016settling} settled the lower bound that rules out PTAS for Nash and matches the QPTAS of \cite{lipton2003playing}, assuming the Exponential Time Hypothesis for PPAD.}

Last but not least, we turn to the learnability of $\delta$-RSE in a setting where the payoff functions are not known in advance but need to be learned from samples of the players' utilities. Such a learning paradigm is crucial to today's common practice of ``centralized
training, decentralized execution'' in multi-agent learning \citep{lowe2017multi,bai2021sample}.  We obtain almost tight results on the learnability of $\delta$-RSE. Specifically, we construct a learning algorithm that, with high probability, outputs a strategy with leader's utility $O(\epsilon)$ or $O(1)$ away from the $\delta$-RSE by using $O(1/\epsilon^2)$ samples, depending on whether a continuity condition is satisfied or not. We then present hard instances with sample complexity lower bounds for each case. As a corollary of this learnability result and the continuity property mentioned, we immediately obtain an algorithm for learning SSE.  This algorithm strictly improves a recent learning algorithm for SSE by  \cite{bai2021sample} on both utility guarantee and computational efficiency.   

\subsection{Related Work}
Stackelberg games have a wide range of applications in economics, finance and security \citep{von2010leadership,van2010dynamic,roth2016watch,kiekintveld2009computing,paruchuri2008playing,tambe2011security}. 
These previous works considered the equilibrium concept, often known as the \emph{strong Stackelberg equilibrium}.
The theory of computing the SSE starts from the seminal work by \citet{conitzer2006computing} and has led to a series of algorithmic studies on Stackelberg games \citep{blum2019computing,conitzer2011commitment,korzhyk2011stackelberg,letchford2010computing}. 
These computation problems of SSE are fundamental to the field of bi-level optimization~\cite{dempe2002foundations} and are viewed as extensions of the classic minimax optimization problem~\cite{v1928theorie} to general sum games.  
A recent work by \citet{goktas2021convex} studied Stackelberg games with dependent strategy sets, based on the constrained maximin model~\citep{wald1945statistical}, and designed first-order methods to efficiently solve the Stackelberg equilibrium under the condition of convex-concave utility and constraints. It turns out that the RSE problem is equivalent to solving a constrained maximin problem where the follower's strategy set is determined by the leader's strategy according to the $\delta$-best response set function. However, as the $\delta$-best response set function exhibits discontinuity without the convex-concave structure, computing RSEs is shown by our Theorem~\ref{thm:hardness} to be computationally intractable.

The robust design problem of Stackelberg strategies has been studied in many recent works, in contexts with uncertain follower utilities  
\citep{letchford2009learning,kiekintveld2013security,kroer2018robust} or follower's uncertain perception of leader's mixed strategies \citep{an2012security,muthukumar2019robust}. 
Another approach is to explicitly model the follower's suboptimal decision-making process with {probabilistic modeling}, such as the well-known {quantal response} model~\citep{mckelvey1995quantal,yang2012computing}.  
The solution concept of RSE is different from the previous robustness notion in that the RSE does not make any specific assumptions about the underlying cause of suboptimal follower behavior. Thus, it is applicable to a wider range of scenarios to address suboptimal follower behavior. 
To our knowledge, \cite{pita2010robust} is the only work that studied a very close RSE solution concept as us. Nevertheless, they focused mostly on experimentally verifying the performance of algorithms based on mixed-integer linear programming for computing robust solutions.

The utility uncertainties have also been considered in machine learning contexts. These results take a learning-theoretic approach and show efficient algorithms to learn the strong Stackelberg equilibrium. One line of research considers the case where the leader is only able to observe follower's responses but not the payoffs~\citep{balcan2015commitment,blum2014learning,letchford2009learning,peng2019learning}. All these works focused on a setting where the follower always optimally responds to the leader's strategy, whereas the follower in our model may respond with any approximately optimal action. 
\rev{Meanwhile, \citet{bai2021sample} studied a setting, where the learner can query any action profile and directly observe bandit feedback of the follower's payoffs. They present a learning algorithm that approximates the strong Stackelberg Equilibria.} We analyze the learnability of the RSE in the same learning setup and present strengthened results as a side product of our result for learning the RSE. 
Our work reveals the intrinsic connections between the robust solution concept and learnability, which in some sense echoes the reduction result from online learning regret to robust mechanism design \citep{camara2020mechanisms}.

More generally, our work subscribes to the rich literature on robust game theory~\citep{aghassi2006robust,bielefeld1988reexamination,camerer2011behavioral,lin2008negotiating,tijs1981nash,tan1995existence, crespi2017robust}. 
In simultaneous games, \cite{tijs1981nash,tan1995existence} studied the existence of $\delta$-equilibrium point as an extension of the Nash equilibrium, where each player's strategy, given other players' strategy profile, has suboptimality at most $\delta$.
Meanwhile, \citet{aghassi2006robust} proposed another equilibrium concept, known as the robust-optimization equilibrium, under which each player, given other players' strategy profile, takes the strategy that maximizes her worst-case utility under the uncertainty of her utility function. This concept is a distribution-free solution concept in contrast to the ex-post equilibrium~\citep{cremer1985optimal} in Bayesian games~\citep{harsanyi1967games}. Among others, \citet{aghassi2006robust, crespi2017robust, perchet2020finding} studied the existence and computation of robust-optimization equilibrium as well as its sensitivity to the robustness parameters, through its connection with the Nash equilibrium of a nominal game (based on worst case payoff function of the original game). Our paper also considers a robust solution concept based on the worst-case optimization in Stackelberg games and studies similar aspects of this solution concept. However, a critical difference between RSE and the robust-optimization equilibrium \citep{aghassi2006robust} is that the source of uncertainty in RSE primarily comes from the opponent's (possibly suboptimal) responses, instead of each player's utility function --- as shown in Section~\ref{sec:learnability}, it is relatively straightforward to accommodate the latter type of uncertainty assuming the continuity of utility over the robustness parameters in our setup.

\section{Preliminaries}\label{sec:prelim}
\paragraph{Stackelberg Games.}
Throughout this paper, we focus on the normal-form Stackelberg game between two players, who are referred to as the leader and the follower, respectively. The leader has the first-mover advantage and commitment power.  
For a Stackelberg game instance $(u_l, u_f)$, $u_l, u_f\in \mathbb{R}^{m \times n}$ denote the leader's and follower's utility matrix, where $m$ (resp. $n$) is the number of leader's (resp. follower's) actions. As a convention in bimatrix games, the $(i,j)$ entry of utility matrix $u_l(i,j)$ (resp. $u_f(i,j)$) is the leader's (resp. follower's) payoff under the action profile $(i,j)$. We also use the standard notation $[m]:= \{1,\dots, m\}$ for the set of $m$ actions and $\Delta^{m} := \{\bvec{x} | \sum_{i=1}^{m} x_i = 1, x_i \geq 0, \forall i\in [m]\}$ for the $m$-dimensional simplex. 

The game has two stages. The leader moves first by committing to a {\em mixed strategy}, $\bvec{x} = (x_1, \cdots, x_m) \in \Delta^{m}$, where each $x_i$ represents the probability the leader playing action $i$. 
The follower then responds to the leader's committed strategy $\bvec{x}$ with some action $j \in [n]$.\footnote{Restricting the follower's response to the pure action set $[n]$ is without loss of generality for our analysis, because in both RSE and SSE, fixing to any leader strategy, even if the follower's optimization problem is relaxed to the mixed strategy space $\Delta^n$, it will always admit vertex solutions. }
At the end, the leader and follower receive the payoff $u_l(\bvec{x}, j), u_f(\bvec{x}, j)$, respectively, where  $u_l(\bvec{x}, j) := \sum_{i\in[m]} x_i \cdot u_l(i, j)$ (and similarly $u_f(\bvec{x}, j)$) is the expected payoff over the randomness of $\bvec{x}$.

\paragraph{A Remark on Approximation.} This paper works extensively with approximate solutions for both computation and learning parts. To be consistent, all approximated solutions we provide are in an additive sense, unless otherwise clarified. As a convention for studying additive approximation, we normalize the entries in all players' utility matrices $u_f, u_l$ to be within the interval $[0, 1]$. This is without loss of generality since rescaling and shifting the utilities will not change a game's equilibrium. 

\paragraph{Strong Stackelberg Equilibrium.}
The conventional solution concept of Stackelberg games is the Strong Stackelberg Equilibrium (SSE), in which the follower always optimally responds to the leader's strategy and breaks ties in favor of the leader, when there is more than one action that maximizes the follower's utility. 
While we have already sketched SSE as the solution to the bilevel optimization program~\eqref{eq:SSE}, it is more convenient to define SSE based on the notion of follower's best response set function as follows.

\begin{definition}[Strong Stackelberg Equilibrium]\label{def:SSE}
In a Stackelberg game $(u_l, u_f)$, we say a strategy profile $(\bvec{x}^*, j^*)$ is a strong Stackelberg equilibrium if it holds that:
\begin{equation}
\label{eq:stackelberg-equilibrium}
         \bvec{x}^* \in \argmax_{\bvec{x} \in \Delta^{m}} \max_{j \in \BR(\bvec{x})} u_l(\bvec{x}, j) 
       \quad \text{ {and} } \quad 
       j^* \in \argmax_{j\in \BR(\bvec{x}^*)} u_l(\bvec{x}^*, j).
\end{equation}
where $\BR(\bvec{x}) := \{j \in [n] | u_f(\bvec{x}, j) \geq \max_{j'\in [n]} u_f(\bvec{x}, j')\}$ denotes the follower's best response set under the leader strategy $\bvec{x}$.
\end{definition}

\paragraph{Robust Stackelberg Equilibrium.}
The best response set function $\BR(\cdot)$ defines an ideal situation, where the follower can always identify her optimal response(s) without any error. In practice, the follower may pick suboptimal responses due to various reasons, such as bounded rationality and limited observations \cite{pita2010robust}.
A natural extension of the best response set, therefore, allows a small error $\delta>0$ in the follower's choice of actions; we define the $\delta$-optimal response set of the follower as
\begin{equation}
\label{def:delta-robust-set}
    \BR_{\delta}(\bvec{x}) := \big\{j\in [n]  \,  \big| \, u_f(\bvec{x}, j) > \max_{j' \in [n]} u_f(\bvec{x}, j')  - \delta  \big\}
\end{equation}
for any leader strategy $\bvec{x} \in \Delta^m$ and $\delta>0$. For the completeness of definition, we denote $\BR_{\delta}(\bvec{x}) = \BR(\bvec{x})$ when $\delta = 0$.\footnote{As we will discuss below, this definition of $\delta$-optimal response set extends the standard best response set in the sense that $\lim_{\delta\to 0^+} \BR_{\delta}(\bvec{x}) = \BR(\bvec{x}) $ under some non-degeneracy assumption.}
As such, we can rewrite the optimization program~\eqref{eq:RSE} into a more convenient definition of the $\delta$-robust Stackelberg equilibrium ($\delta$-RSE) as follows.

\begin{definition}[$\delta$-Robust Stackelberg Equilibrium]\label{def:delta-robust-SSE} 
In a Stackelberg game $(u_l, u_f)$, for any $\delta>0$, we say a strategy profile $(\bvec{x}^*, j^*)$ is a $\delta$-robust Stackelberg equilibrium if it holds that:
\begin{equation}
\label{eq:epsilon-robust-equilibrium}
         \bvec{x}^* \in \argmax_{\bvec{x} \in \Delta^{m}} \min_{j \in \BR_{\delta}(\bvec{x})}  u_l(\bvec{x}, j) 
        \quad \text{ { and} }\quad
        j^* \in \argmin_{j\in \BR_{\delta}(\bvec{x}^*)} u_l(\bvec{x}^*, j).
\end{equation}
In addition, we denote $u_{\textup{RSE}}(\delta) := u_l(\bvec{x}^*, j^*)$ as the leader's utility obtained in a $\delta$-RSE.
\end{definition}

\rev{It is worth noting that there are other potential definitions of the game equilibrium, too, e.g., the leader's strategy may be restricted to a pure one, while the follower may play a mixed strategy instead. 
We choose to focus on the definition presented above over other alternatives for the following reasons.
First, from a modeling perspective, Stackelberg games differ from simultaneous-move games by explicitly modeling information and strategic asymmetries between players. In most studies of Stackelberg games, including the seminal work of \citet{von2010leadership} as well as perhaps its most successful modern application to security games \cite{tambe2011security}, the leader is assumed to play a mixed strategy, whereas the follower observes the leader’s mixed strategy and then best responds with a pure strategy.\footnote{Similar to simultaneous-move games (e.g., rock-paper-scissor), the motivation of mixed strategies also arises from repeated interactions during which the follow may observe past realizations of the strategy which shapes/confirms the follower's beliefs about leader's commitment.} Our work follows this standard modeling convention, as the robust Stackelberg equilibrium assumes that the leader anticipates a range of possible follower responses and prepares for the worst-case among them.

Second, from a computational perspective, allowing the follower to choose any mixed strategy supported on the current $\delta$-best response set, as defined in Eq. \eqref{def:delta-robust-set}, would not change our analysis: the follower’s worst-case response still corresponds to a vertex solution, which is a pure strategy in the $\delta$-best response set.
One could consider yet another definition of the $\delta$-best response set that includes all $\delta$-suboptimal mixed strategies. Though mathematically natural, this definition lacks strong practical motivations, since in reality it is more common that the follower experiences uncertainty or difficulty in identifying an optimal pure response, rather than deliberately selects a worst-case mixed strategy that could put positive probabilities on obviously bad follower responses.
Therefore, we do not consider this alternative definition in this paper.
}

\paragraph{Inducibility Gap and Non-Degeneracy.}
We also introduce a notion called the \textit{inducibility gap}, denoted by $\Delta$. This quantity relates to the hardness of robustifying Stackelberg leader strategy, and it plays an important role in our analysis of $\delta$-RSE in this paper.

\begin{definition}[Inducibility Gap]\label{def:delta_inducable}
In a Stackelberg game $(u_l, u_f)$, the inducibility gap is the largest constant $\Delta$ such that for any follower actions $j \in [n]$, there exists a leader strategy $\bvec{x}^j$ with 
$
u_f(\bvec{x}^j, j) \geq \max_{j' \neq j} u_f(\bvec{x}^j, j') + \Delta.
$
\end{definition} 

Namely, the inducibility gap is an intrinsic property of the game on how easy it is for the leader to incentivize the follower to play any action. More specifically, let $\Delta(\bvec{x}, j) :=  u_f(\bvec{x}, j) - \max_{j'\neq j} u_f(\bvec{x}, j') $ denote the follower's utility margin of taking action $j$ under leader's strategy $\bvec{x}$ --- a  measure of how inducible is $j$ under $\bvec{x}$ (e.g., if $\Delta(\bvec{x}, j) > \delta$, then $\BR_{\delta}(\bvec{x}) = \{j\}$).
Then, we can alternatively define the inducibility gap through the ``minimax'' lens:
$$ \Delta = \max_{\bvec{x} \in \Delta^m} \min_{j\in [n]} \Delta(\bvec{x}, j). $$

We remark that under the classic solution concept of SSE, it is without loss of generality to assume $\Delta > 0$. First, any generic game has $\Delta \neq 0$, because for randomly generated game instances, the events that two actions always have the same payoff have zero measure~\citep{von2010leadership}.
Moreover, if a game has $\Delta < 0$, then there exists some follower action $j$ that can never be the best follower response (thus will not be played): more formally through contraposition of Definition \ref{def:delta_inducable}, there exists follower action $j$ such that for any $\bvec{x}$ we have 
$u_f(\bvec{x}, j) <  u_f(\bvec{x}, j') + \Delta < u_f(\bvec{x}, j')
$ for some $j' \in [n]$. Therefore, it is without loss of generality to ignore and remove this never-to-be-played follower action $j$ (regardless of solving or learning the game); consequently, we obtain a Stackelberg game in which every follower action at least can possibly be a best response, and such a game has $\Delta > 0$. However, for $\delta$-RSE, it is only without loss of generality to
assume $ \Delta > -\delta$, since $\delta$-suboptimal follower actions will affect the $\delta$-RSE.
Some of our results about $\delta$-RSE hold under assumption $\Delta > 0$, which slightly
 loses generality compared to $\Delta > -\delta$.  This discrepancy
becomes smaller as $\delta$ decreases (i.e., the follower becomes less suboptimal). Overall we believe it is a reasonable assumption to pursue when $\delta$ is small.

We conclude this section with an example of Stackelberg game motivated by the real world application and illustrate the both solution concepts of SSE and $\delta$-RSE.  
\begin{example}[Stackelberg Competition]
The Stackelberg game origins from the duopoly competition model studied by \citet{von2010market}. There are two firms who are selling the same type of product. The leader firm is able to enter a market earlier than the follower firm. The leader firm has two options, to set either a high or low price for the product. The follower firm also have two options, which is to either compete with the leader or leave this market. 
Their payoff matrices can be formulate as follows. 
\begin{center}
     \begin{tabular}{ | l|  c|  c|}  %
        \hline
       \ $u_{l}, u_{f}$ & $j_1$ (compete) & $j_2$ (leave)\\
        \hline
    $i_1$ (high) & $3, 2$ & $6,1$\\ 
        \hline
    $i_2$ (low) & $2, 0$ & $4,1$\\
        \hline
    \end{tabular}   
\end{center}

Under this pair of payoff matrices, we can observe that if the leader prices the product at a high price, then the follower is willing to compete. Otherwise, the follower would choose to leave the market. To design the Stackelberg strategy, the leader has the power to randomize her actions (e.g., to offer a price in the middle of high and low). Suppose the follower chooses the best response and break tie optimistically, we can plot the leader's utility function $\max_{j\in \BR(\bvec{x})} u(\bvec{x}, j)$ by the red shaded line and its maximum gives the SSE leader strategy, $\bvec{x}^* = (\frac{1}{2}, \frac{1}{2})$ with the follower response $j_1$.
For various reasons in practice, the follower could choose any action from his $\delta$-best response set and break tie pessimistically. In this case, the leader's utility function becomes $\min_{j\in \BR_{\delta}(\bvec{x})} u(\bvec{x}, j)$, plotted as the green shaded line and its maximum gives the $\delta$-RSE leader strategy, $\bvec{x}^* = (\frac{1+\delta}{2}, \frac{1-\delta}{2})$ with the follower response $j_1$.
\begin{figure}[h]
    \centering
    \includegraphics[width=0.4\textwidth]{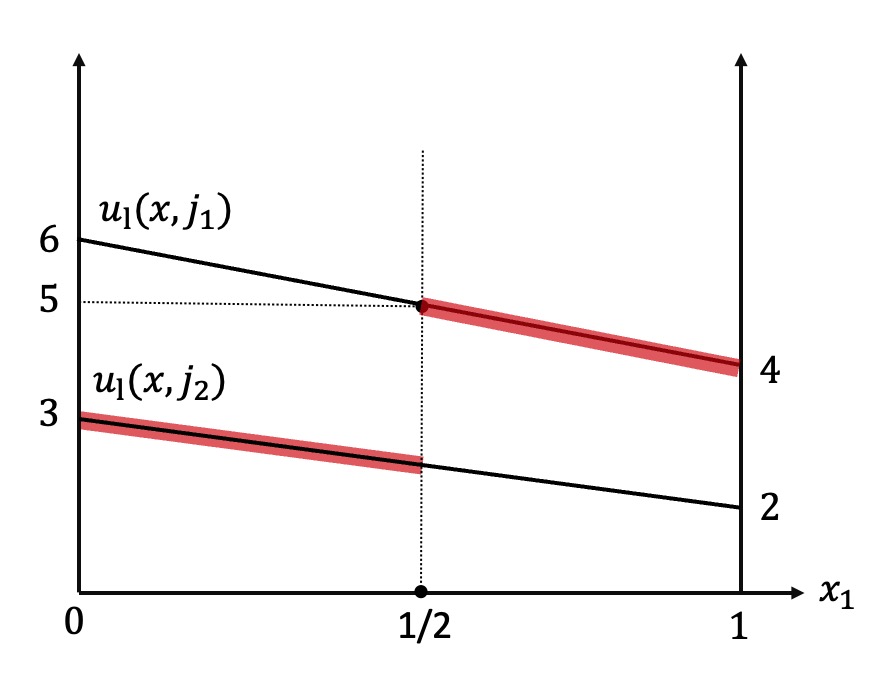}
    \includegraphics[width=0.4\textwidth]{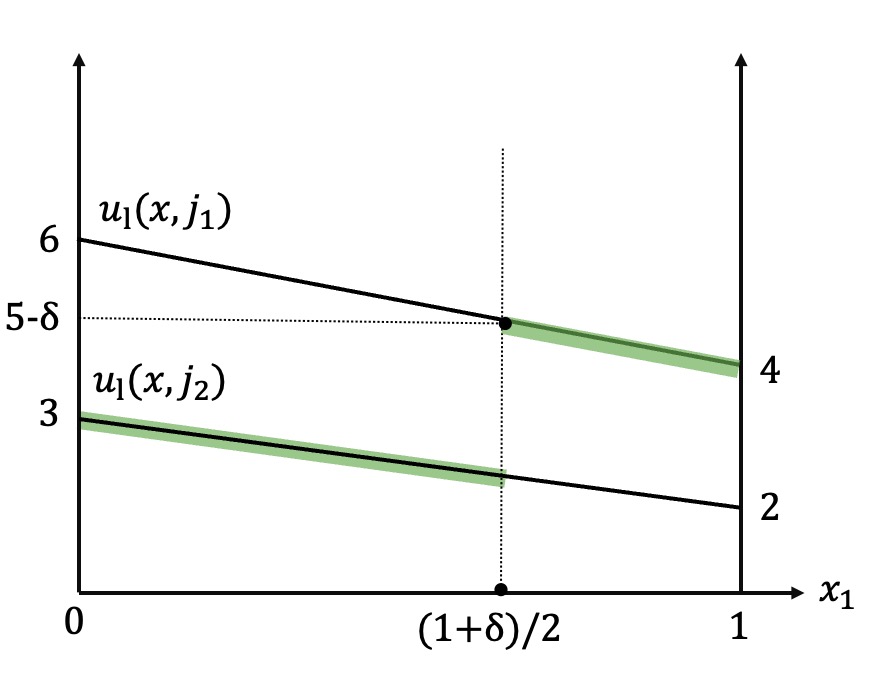}
    \caption{Plots of leader utility functions under follower response models in SSE (left) and $\delta$-RSE (right). The x-axis is the probability of the leader choosing $i_1$, while the y-axis is the leader payoff. }
    \label{fig:stackelberg-competition}
\end{figure}
\end{example}

\section{Analytic Properties of RSE}\label{sec:RSE-Property}

\subsection{On the Alternative Definitions of $\delta$-RSE} 
In this section, we show how our current definition of $\delta$-RSE is an inevitable choice by comparing it with several plausible alternative choices.
The first objection one may raise is that the notion of $\delta$-RSE may not be well-defined, in the sense that $\min_{j \in \BR_{\delta}(\bvec{x})}  u_l(\bvec{x}, j)$ (as a function of $\bvec{x}$) may not always have a maximum.
This has been a concern for some variants of the Stackelberg equilibrium such as the {\em weak Stackelberg equilibrium}, where the follower is assumed to break tie pessimistically~\citep{von2010leadership}. 
It turns out that $\delta$-RSE is indeed well-defined. Here, the strict inequality in Equation~\eqref{def:delta-robust-set} is crucial. We present a simple example below to illustrate the subtlety involved in the definition: a $\delta$-RSE may not exist if responses that are exactly $\delta$-optimal are included $\BR_{\delta}(\bvec{x})$ in Equation~\eqref{def:delta-robust-set}.

\begin{example}[Non-existence of $\delta$-RSE under an Alternative Definition]\label{ex:non-existence}
We claim that in the following game instance, no $\delta$-RSE exists under a slightly different definition of $\delta$-optimal response set: $\BR_{\delta}(\bvec{x}) = \{j | u_f(\bvec{x}, j) \geq u_f(\bvec{x}, j') - \delta, \forall j' \neq j\}$. 
\begin{center}
     \begin{tabular}{|c|  c|  c|}  %
        \hline
        $u_{l}, u_{f}$ & $j_1$ & $j_2$ \\
        \hline
        $i_1$ & $0, -1$ & $0, 0$\\ 
        \hline
        $i_2$ & $0, -\delta$ & $1,0$\\
        \hline
        $i_3$ & $0, 1$ & $0,0$\\
        \hline
    \end{tabular}   
\end{center}
Notice that the leader's optimal strategy should ensure that  $j_1 \not\in \BR_\delta(\bvec{x})$, because the leader's utility would otherwise always be zero due to the follower's pessimistic tie-breaking. This means that $i_3$ can be dropped, since it only encourages the follower to take $j_1$ while gives the leader zero utility. Hence, it suffices to set the leader strategy as $\bvec{x} = (\xi, 1-\xi, 0)$, under which the follower's $\delta$-optimal response set (under the modified definition) would be $\BR_\delta(\bvec{x}) = \{j_2\}$, resulting leader utility $u_l(\bvec{x}, j_2) = 1-\xi$. The smaller the $\xi$ is, the larger the leader utility is. It is now clear that the optimal leader strategy is to play $i_1$ with infinitely small probability $\xi > 0$ while play $i_2$ with probability $1-\xi$. 
However, $\xi$ can't be equal to $0$, as this would include $j_1$ into the $\delta$-best follower response set that makes the leader's utility become $0$. 
Therefore, since $\xi$ needs to be infinitely small but greater than $0$, the $\delta$-RSE is not well defined and does not exist for this example. 
\end{example}

Example \ref{ex:non-existence} shows that the leader may optimize over an \emph{open} set of mixed strategies that induces the desirable follower response behaviors and thus cannot achieve the exact optimal strategy, but it is unclear how our definition of $\BR_{\delta}(\cdot)$ in Equation \eqref{def:delta-robust-set} can avoid this problem.
Below, we present a constructive proof to show that the $\delta$-RSE we defined via Equation \eqref{def:delta-robust-set} and \eqref{eq:epsilon-robust-equilibrium}  above  exists  in every game for any $\delta > 0$.

\begin{proposition}\label{thm:existence}
A $\delta$-RSE under Definition \ref{def:delta-robust-SSE} exists in every game for any $\delta > 0$ and can be computed in $O(2^n \poly(m,n))$ time.
\end{proposition}
\rev{
The proof idea of Proposition \ref{thm:existence} is to explicitly provides an algorithm for computing a $\delta$-RSE, though it runs in exponential time in the worst case. %

\textit{Proof.}\quad
We provide a constructive proof for the theorem by exhibiting an algorithm that computes a $\delta$-RSE for any Stackelberg game. The main idea of our algorithm is to partition the simplex in sub-regions and search for a $\delta$-RSE candidate within each sub-region $\mathcal{X}_{(S,  \tilde j, j)}$ by solving a linear program but with \emph{relaxed} non-strict inequalities (thus in total we solve exponentially many LPs). The crux of our proof is to argue that, while in general the relaxation above is not tight,  there always exists a region $\mathcal{X}_{(S,  \tilde j, j)}$ for which the above relaxation is tight and moreover the leader achieves the best possible leader utility. This proves the existence of a $\delta$-RSE.

We begin with the definition and analysis of different sub-regions of the leader's strategy space $\Delta^m$. Each sub-region $\mathcal{X}_{(S,  \tilde j, j)}$ is  characterized by three factors: (1) Follower $\delta$-optimal response action set $S \in 2^{[n]}$; (2) Follower action $\tilde{j} \in S$ with maximal follower utility among actions in $S$; (3) Follower action $j \in S$ with the worst leader utility among actions in $S$ ($j$ and $\tilde{j}$ are different in general). Mathematically, for any follower action set $S \in 2^{[n]}$ and any $\tilde{j}, j \in S$,
\begin{small}\begin{multline*}\label{eq:leaderStrategySubregion}
\mathcal{X}_{(S,  \tilde j, j)} := \bigg \{\bvec{x} \in \Delta^m \,  \bigg| \,  \BR_{\delta}(\bvec{x}) = S, \, \, u_f(\bvec{x}, \tilde j) \geq u_f(\bvec{x}, k),   u_l(\bvec{x}, j) \leq u_l(\bvec{x}, k), \forall k \in S  \bigg \}. 
\end{multline*} \end{small}
Next, we make a few observations about $\mathcal{X}_{(S,  \tilde j, j)}$. 

First, $\bigcup_{\tilde j, j \in S, S\in 2^{[n]}} \mathcal{X}_{(S,  \tilde j, j)}  = \Delta^m$. This is because any $\bvec{x} \in \Delta^m$ must induce some follower $\delta$-optimal response set $S$, an optimal follower action as $\tilde j$, and a follower response $j$ that is the worst response for the leader. As a result, any $\bvec{x}$ must belong to some $\mathcal{X}_{(S,  \tilde j, j)}$.

Second, $\mathcal{X}_{(S,  \tilde j, j)}$ is a (possibly open) \textit{polytope} that can be expressed by linear constraints, with \textit{strict} inequality constraints. This is because both constraints $ u_f(\bvec{x}, \tilde j) \geq u_f(\bvec{x}, k) $ and $  u_l(\bvec{x}, j) \leq u_l(\bvec{x}, k) $ are linear in variable $\bvec{x} = (x_1,\cdots,x_m)$. Moreover, we have the following equivalent expression for the constraint $ \BR_{\delta}(\bvec{x}) = S$:
\begin{equation}\label{eq:robust-response-inequalities}
  \BR_{\delta}(\bvec{x}) = S \qquad \Longleftrightarrow  \qquad  \begin{cases}
u_f(\bvec{x}, k) > u_f(\bvec{x}, \tilde j) - \delta ,  \,   \,  \forall k \in S,  \qquad \text{ and} \\
u_f(\bvec{x}, k) \leq  u_f(\bvec{x}, \tilde j) - \delta ,  \,   \,  \forall k \not \in S. 
\end{cases}  
\end{equation}
Notably, the second set of constraints guarantees that any $k \notin S$ cannot be a $\delta$-optimal  follower response, which is essential for the definition of $S$. Different from typical linear constraints, the first constraint above is a strict inequality and thus the resultant polytope $\mathcal{X}_{(S,  \tilde j, j)}$ might be an open set. 

In search of the optimal leader strategy, the main idea of our algorithm is to solve multiple (in fact, exponentially many) linear programs by relaxing the strict inequality in \eqref{eq:robust-response-inequalities}. The correctness proof of  this algorithm  relies on a key insight that there always exists a tuple $(S, \tilde j, j)$ such that the leader objective on this tuple  achieves the best possible leader utility and moreover the above relaxation of the $\delta$-optimal  follower response constraints will be ``tight'' on this particular tuple (although  this relaxation is not tight for other tuples in general). 

\begin{algorithm}[tbh]
    \caption{\textsc{Computing $\delta$-RSE via LP Relaxations}}
    \label{alg:delta-RSE-Alg}
\SetKwInOut{Input}{Input}
\SetKwInOut{Output}{Output}

\Input{leader utilities $u_l \in \RR^{m\times n}$, follower utilities $u_f \in \RR^{m\times n}$, parameter $\delta > 0$.}

\Output{$\delta$-RSE $(\bvec{x}^*, j^*)$.}
    
\BlankLine

\For{any non-empty $S \subseteq [n]$ and any $\tilde j, j \in S$} 
{
Solve the following \emph{relaxed} linear program, denoted by $LP(S, \tilde j, j)$,  to compute an optimal leader strategy $\bvec{x}^*_{S,\tilde j, j}$ within (a relaxation of) $\mathcal{X}_{(S,  \tilde j, j)}$:
\begin{lp}
\label{lp:delta-RSE-LP}
    \maxi{ \sum_{i=1}^m x_i \cdot u_l(i,j) }
    \st 
    \qcon{ \sum_{i=1}^m x_i \cdot u_f(i,\tilde{j}) \geq   \sum_{i=1}^m x_i u_f(i,k)  }{k  \in [n] } 
    \qcon{ \sum_{i=1}^m x_i \cdot u_f(i,k) \geq   \sum_{i=1}^m x_i \cdot u_f(i,\tilde j) - \delta     }{k \in S \quad \text{(relaxed constraints)}}
    \qcon{ \sum_{i=1}^m x_i \cdot u_f(i,k) \leq  \sum_{i=1}^m x_i \cdot u_f(i,\tilde j) - \delta     }{k  \not \in S}
    \qcon{ \sum_{i=1}^m x_i \cdot u_l(i,j) \leq    \sum_{i=1}^m x_i \cdot u_l(i,k)     }{k \in S } 
    \con{ (x_1,\cdots,x_m)\in \Delta^m}
\end{lp}
}

Choose any tuple $(S^*,  \tilde j^*, j^*)$ that has the maximum leader utility. That is,  let $\bvec{x}^* = \bvec{x}^*_{S^*,\tilde j^*, j^*}$ then   $u_l(\bvec{x}^*, j^*) \geq u_l(\bvec{x}^*_{S, \tilde j, j}, j)$ for any $S \subseteq [n]$ and any $\tilde j, j \in S$.  

Verify the second constraint of LP \eqref{lp:delta-RSE-LP} for the chosen tuple $(S^*,  \tilde j^*, j^*)$,  and let $\hat{S}$ denote the set of all $k \in S^*$ such that the second constraint is \emph{strict} at $\bvec{x}^*$ for the $k$. 

Let $\hat j \in \hat S$ denote the follower action with minimum leader utility, i.e., $u_l (\bvec{x}^*, \hat j) \leq u_l (\bvec{x}^*, k) $ for any $k \in \hat S$.

\Return $(\bvec{x}^*, \hat j)$.
\end{algorithm}

For the convenience of exposure, we present the formal procedure in Algorithm \ref{alg:delta-RSE-Alg}. At the high level, the algorithm enumerates all possible tuple $(S, \tilde j, j)$ (Line 1),  solves LP \eqref{lp:delta-RSE-LP} for the optimal leader strategy $\bvec{x}^*_{S,\tilde j, j}$ within (a relaxation of) the polytope $\mathcal{X}_{(S,  \tilde j, j)}$ (Line 2), and finally picks the $LP(S^*,  \tilde j^*, j^*)$ whose solution $\bvec{x}^*$ has the highest leader utility (Line 4). Notably, in order to optimize over a closed polytope to avoid the non-existence of optimal solutions, we had to relax the $\delta$-optimal response constraint, i.e., the second constraint in LP \eqref{lp:delta-RSE-LP}, to be ``$\geq$''. Therefore, we thus suffer the risk of getting a solution $\bvec{x}^*$ that cannot truly induce the expected follower $\delta$-optimal response set $S^*$, i.e., those actions $k \in S^*$ such that  $u_f(\bvec{x}^*, k) = u_f(\bvec{x}^*, \tilde j^*) - \delta $. Here comes the crucial (though seemingly straightforward) step in Lines 5 and 6 to remove the invalid  $\delta$-optimal responses from the set $S^*$ and ultimately construct a valid equilibrium strategy.

It is easy to see that, with the adjustment  at Line 5 and 6, Algorithm \ref{alg:delta-RSE-Alg} always outputs a valid strategy pair  $(\bvec{x}^*, \hat j)$ in the sense that $\hat j \in \BR_{\delta}(\bvec{x}^*)$ and  $\hat j$ is indeed the follower response that is worst for the leader in the $\delta$-optimal response set under leader strategy  $\bvec{x}^*$. This follows simply by the construction of the algorithm, which removes all invalid follower
$\delta$-optimal responses $k \in S^*$, due to $u_f(\bvec{x}^*, k) =   u_f(\bvec{x}^*, \tilde j^*) - \delta $, out the set $S^*$. This adjustment leads to a strictly small subset $\hat S$, which is truly the follower $\delta$-optimal response set $\BR_{\delta}(\bvec{x}^*)$. It is worth noting that this $\hat S$ is always non-empty for any $\delta > 0$, since $u_f(\bvec{x}^*, \tilde j^*) > u_f(\bvec{x}^*, \tilde j^*) - \delta$. Therefore, we have $\tilde j^* \in \hat S$ for any $\delta >0$. The algorithm then ``re-appoints'' $\hat j \in \hat S$ as the true worst follower response within $\hat S$ in the sense of minimizing the leader utility.  

What remains to argue is that the output strategy $(\bvec{x}^*, \hat j)$ achieves at least the leader utility as that of a $\delta$-RSE, and thus must be a $\delta$-RSE strategy pair.   First, since LP \eqref{lp:delta-RSE-LP} is a relaxation of the $\delta$-RSE problem and $\cup_{\tilde j, j \in S, S\in 2^{[n]}} \mathcal{X}_{(S,  \tilde j, j)}  = \Delta^m$, the optimal objective value of  LP \eqref{lp:delta-RSE-LP} for the special tuple  $(S^*,  \tilde j^*, j^*)$ picked at Line 4 of Algorithm \ref{alg:delta-RSE-Alg} must be as large as the leader's $\delta$-RSE utility. Therefore, all we need to argue is that the leader utility under the valid strategy pair $(\bvec{x}^*, \hat j)$ is at least the optimal objective of $LP(S^*,  \tilde j^*, j^*)$, which is exactly  $u_l (\bvec{x}^*, j^*)$.  This is true because both $\hat{j}, j^* \in S^*$, and the fourth constraint of $LP(S^*,\tilde j^*, j^*)$ --- i.e., $j^*$ has the worst leader utility among all actions in $S^*$ --- implies  $ u_l (\bvec{x}^*, j^*) \leq u_l (\bvec{x}^*, \hat{j})$ which is precisely the leader utility for the feasible strategy pair  $(\bvec{x}^*, \hat j)$.  

Lastly, for the time complexity,
Algorithm~\ref{alg:delta-RSE-Alg} solves $O(n^2 2^n)$ linear programs of size $O(mn)$, each corresponding to a tuple $(S, \tilde j, j)$. Hence, the stated time complexity follows.
\qed}

We make a few remarks about Proposition  \ref{thm:existence}. First, it implies that for instances with a small $n$, a $\delta$-RSE can be computed efficiently.
Nevertheless, the exponential dependency on $n$ in the time complexity appears to be inevitable, as we will show in the next section that computing a $\delta$-RSE is NP-hard. 
\rev{Second, one may wonder why approaches similar to Algorithm~\ref{alg:delta-RSE-Alg} cannot be used to prove the existence of a $\delta$-RSE under the alternative definition with $\geq$ in Equation~\ref{def:delta-robust-set}. 
The reason is that, with this alternative definition, the third constraint in Equation~\eqref{lp:delta-RSE-LP} of Algorithm~\ref{alg:delta-RSE-Alg} will become a strict inequality.
Consequently, the set $\hat{S}$ will include additional follower actions that would have been excluded under our current $\delta$-RSE definition.
This will then lead to $\hat{S} \supseteq S^*$ rather than $\hat{S} \subseteq S^*$, making the remaining part of the proof of Proposition~\ref{thm:existence} inapplicable.
}
Third, while one may be willing to compromise and settle with the ``supremum'' of leader strategies, instead of using the exact ``maximum'' in Equation~\eqref{eq:epsilon-robust-equilibrium}, this is generally viewed as being undesirable. Such a wrinkle of the non-existence of a solution concept is always a concern for game-theoretical analysis. 
Hence, we believe it is helpful to figure out the right way so that we do not always need to worry about the existence and how to tweak the solution to make it achievable.\footnote{For example, \citet{pita2010robust} used non-strict inequalities for both the $\delta$-optimal  follower response and non-$\delta$-optimal  follower response. 
This choice actually leads to strange inconsistency in the follower behavior modeling, in which the follower will effectively break ties \emph{against} the leader among actions strictly within $\delta$-optimal response region but then \emph{in favor of} the leader at the boundary of $\delta$-optimal response regions.}  This also paves the way for many of our following analysis.

Another reasonable concern here is that given the existing equilibrium concepts, is it truly necessary to define and study the $\delta$-RSE solution concept? In particular, might it be that the SSE leader strategy or the maximin leader strategy will already  perform well, i.e., achieve $\epsilon$-optimal $u_{\textup{RSE}}(\delta)$ assuming the suboptimal follower   responses? 
Unfortunately, Proposition~\ref{prop:equilibrium_variants} shows that both the SSE leader strategy and maximin leader strategy are highly suboptimal, with a $\Omega(1)$ suboptimality gap compared with a $\delta$-RSE. 
As a result, we cannot simply apply the leader strategy from other equilibrium (e.g. SSE) to expect robust performance on par with $\delta$-RSE. See Appendix \ref{appendix-property-suboptimal} for the proof of Proposition~\ref{prop:equilibrium_variants}.

\begin{proposition}[Suboptimality of Standard Equilibrium Strategies]
\label{prop:equilibrium_variants}
For any $\delta > 0$,
there exists game instances in which both the SSE leader strategy ${\bvec{x}}_1$ and maximin leader strategy ${\bvec{x}}_2$ have a constant suboptimality gap of at least $1/2$ as approximations to a $\delta$-RSE, i.e.,
$$
\min_{j \in \BR_{\delta}({\bvec{x}}_1)} u_l({\bvec{x}}_1, j) < u_{\textup{RSE}}(\delta) - \frac{1}{2} \quad \textup{and} \quad
\min_{j \in \BR_{\delta}({\bvec{x}}_2)} u_l({\bvec{x}}_2, j) < u_{\textup{RSE}}(\delta) - \frac{1}{2}.
$$
\end{proposition} 

The other common question of the $\delta$-RSE definition is regarding the follower's pessimistic tie-breaking behavior. That is, how does the tie-breaking behavior affect the performance of $\delta$-RSE against the suboptimal follower responses?
Similar question is asked about the SSE, and \citet{von2010leadership} showed that the SSE enjoys generically unique leader payoff regardless of the follower's tie-breaking rules. 
Does $\delta$-RSE share similar property? Would the leader payoff under $\delta$-RSE strategy $\bvec{x}^*$ stay the same if the follower could switch to different tie-breaking rules?
The question becomes non-trivial if we were to restrict to reasonably small $\delta$ such that $\delta < \Delta$. 
It turns out that the answer is ``No'', as shown in Proposition~\ref{prop:tie-breaking} below. This result confirms the necessity of our worst case analysis under pessimistic tie-breaking.

\begin{proposition}[Tie-breaking Rule Matters for $\delta$-RSE] \label{prop:tie-breaking}
For any $\delta > 0$,
there exist game instances with inducibility gap $\Delta > \delta$ in which the leader payoffs of $\delta$-RSE leader strategy $\bvec{x}^*$ under the optimistic and pessimistic tie-breaking rule have a difference of at least $\delta$, i.e., 
$$ \min_{j\in \BR_{\delta}(\bvec{x}^*) } u_{l}(\bvec{x}^* , j) < \max_{j\in \BR_{\delta}(\bvec{x}^*) } u_{l}(\bvec{x}^* , j) - \delta . $$
\end{proposition}

See Appendix~\ref{append:prop-tie-breaking} for the constructed instance with $m=3$. Meanwhile, we show in Appendix~\ref{sec:other-properties} that the $\delta$-RSE strategy $\bvec{x}^*$ does have unique leader payoff regardless of the follower's tie-breaking rules in any generic game with $m=2$, which leads to another efficient algorithm to compute $\delta$-RSE when $m$ is constant.

\subsection{{The Price of Robustness: $\delta$-RSE Leader Utility Curve over $\delta$ }}
\begin{figure}[h]
    \centering
    \includegraphics[width=0.4\textwidth]{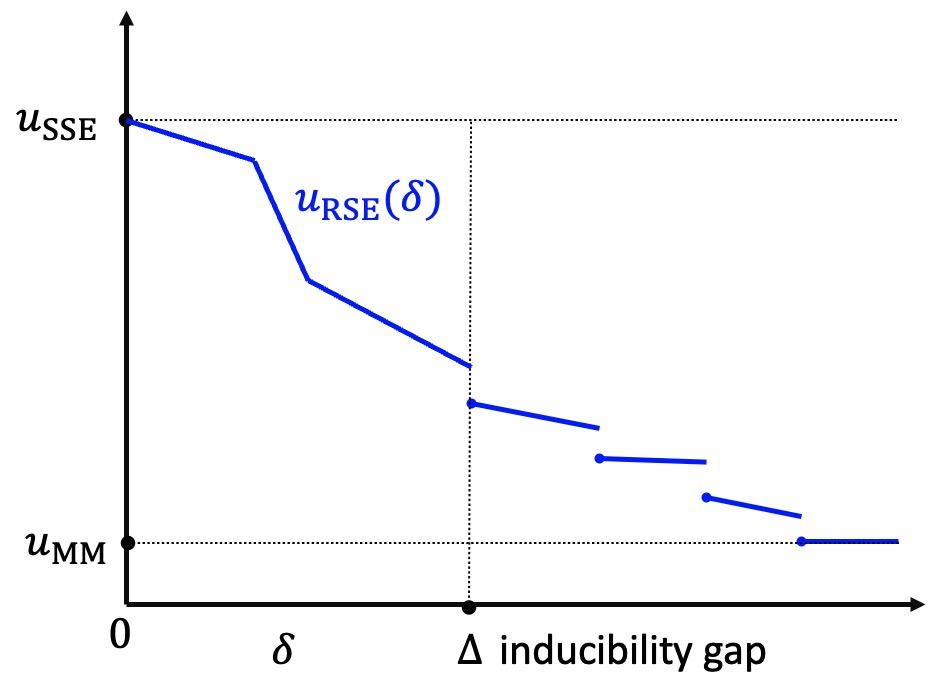}
    \caption{An illustration of $u_{\textup{RSE}}(\delta)$, the leader utility in $\delta$-RSE w.r.t. $\delta > 0$.  }
    \label{fig:rse}
\end{figure}

Given the analysis above, we know the leader's utility in a $\delta$-RSE is a well-defined function in the domain of $\delta>0$, for the existence and uniqueness of its value. 
We denote this function as $u_{\textup{RSE}}(\delta)$. Next, we derive the main results of this section about characteristics of $\delta$-RSE through $u_{\textup{RSE}}(\delta)$. The $\delta$ interval with continuity property is especially crucial for our study of $\delta$-RSE in the following sections.
To provide an intuitive understanding of Theorem~\ref{thm:delta-RSE-property}, we depict a typical shape of the function $u_{\textup{RSE}}(\delta)$ in Figure~\ref{fig:rse}.

\begin{theorem}\label{thm:delta-RSE-property}
Denote the leader's SSE payoff as $u_{\textup{SSE}}$, and the leader's maximin payoff as
$u_{\textup{MM}} = \max_{\bvec{x}\in\Delta^m} \min_{j \in [n]} u_l(\bvec{x}, j)$,
the following properties of $\delta$-RSE hold for any game:
\begin{enumerate}
\item\label{rse-property-1}
For any $\delta, \delta'$ such that $0 < \delta \leq \delta'$, $u_{\textup{RSE}}(\delta)$ is monotone non-increasing and is bounded as,
\begin{equation}\label{eq:property1part1}
u_{\textup{SSE}} \geq  u_{\textup{RSE}}(\delta) \geq u_{\textup{RSE}}(\delta') \geq u_{\textup{MM}},
\end{equation}  
where $u_{\textup{RSE}}(0^+) := \lim_{\delta \to 0^+} u_{\textup{RSE}}(\delta)$ exists; moreover, if  $\Delta >0$,  $u_{\textup{RSE}}(0^+) = u_{\textup{SSE}}$.

\item\label{rse-property-2} 
For any $\delta$ such that $0 < \delta < \Delta$, it holds that
\begin{equation}\label{eq:property1part2}
u_{\textup{RSE}}(\delta) \geq u_{\textup{SSE}} - \delta/\Delta.
\end{equation}

\item\label{rse-property-3}  
For any $\Delta > 0$, $u_{\textup{RSE}}(\delta)$ is $L$-Lipschitz continuous when $\delta \in (0, \Delta - \frac{1}{L}]$ and $L > 1/\Delta$.  Meanwhile, $u_{\textup{RSE}}(\delta)$ can be discontinuous when $\delta \geq \Delta$.

\end{enumerate}
\end{theorem}

Before proceeding to the proof, we make a few remarks about Theorem \ref{thm:delta-RSE-property}. First, Property \ref{rse-property-1} suggests that the equality for $u_{\textup{SSE}} \geq  u_{\textup{RSE}}(\delta)$ can be attained, if the non-degeneracy condition $\Delta >0$ is satisfied. Note that the asumption $\Delta > 0$ is necessary as  in Appendix \ref{sec:appendix-proof-section-property1} we present an instance with $\Delta = 0$ such that $u_{\textup{RSE}}(0^+) < u_{\textup{SSE}}$.  
Property \ref{rse-property-2} shows that with the inducibility gap $\Delta$, the lower bound of the leader's utility under $\delta$-RSE can be improved from $u_{\textup{MM}}$ to be $u_{\textup{SSE}} - \delta/\Delta$. Later in Section \ref{sec:computational} of algorithmic studies, we will show how this property   leads to a simple approximation algorithm for $\delta$-RSE. 
Lastly, Property \ref{rse-property-3} shows that $u_{\textup{RSE}}(\delta)$ is Lipschitz continuous when $\delta < \Delta$. 
This Lipschitz continuity turns out to be very useful for learning a $\delta$-RSE in contexts where the follower utility is not known in advance. We will demonstrate its applicability to learning in Section \ref{sec:learnability}.

\smallskip
\noindent\textit{Proof of Theorem \ref{thm:delta-RSE-property}.}\quad 

\noindent{\bf Property \ref{rse-property-1}.}  
We begin with the monotonicity of $u_{\textup{RSE}}(\delta)$. According to the definition of follower's $\delta$-optimal response set in Equation~\eqref{def:delta-robust-set}, $\BR_{\delta}(\bvec{x})$ expands with $\delta$,
i.e., for any leader strategy $\bvec{x}$, we have
\begin{equation*}\label{eq:delta-response-sets}
    \BR_{\delta}(\bvec{x}) \subseteq \BR_{\delta^{'}}(\bvec{x}), \quad \forall 0 < \delta \leq \delta^{'}.
\end{equation*}
Recall that $u_{\textup{RSE}}(\delta) = \max_{\bvec{x} \in \Delta^{m}} \min_{j \in \BR_{\delta}(\bvec{x})}  u_l(\bvec{x}, j)$. Hence, for any $\bvec{x} \in \Delta^m$, $0<\delta \leq \delta^{'}$, we have
\begin{equation*}\label{eq:delta-vs-delta-prime-utility}
    \begin{split}
        u_{\textup{RSE}}(\delta) = \max_{\bvec{x} \in \Delta^{m}} \min_{j \in \BR_{\delta}(\bvec{x})}  u_l(\bvec{x}, j) 
        \geq \max_{\bvec{x} \in \Delta^{m}} \min_{j \in \BR_{\delta^{'}}(\bvec{x})}  u_l(\bvec{x}, j)
        = u_{\textup{RSE}}(\delta^{'}).
    \end{split}
\end{equation*}
For the lower bound of $u_{\textup{RSE}}(\delta)$, let $\delta \geq \max_{i\in [m], j,j'\in [n]} u(i,j)-u(i,j')$. Notice that we have $\BR_{\delta}(\bvec{x}) = [n]$, for any $\bvec{x}$. This readily implies,
\begin{equation*}\label{eq:delta-vs-max-min-utility}
    \begin{split}
        u_{\textup{RSE}}(\delta) = \max_{\bvec{x} \in \Delta^{m}} \min_{j \in \BR_{\delta}(\bvec{x})}  u_l(\bvec{x}, j) 
        = \max_{\bvec{x} \in \Delta^{m}} \min_{j \in [n]}  u_l(\bvec{x}, j)
        = u_{\textup{MM}}.
    \end{split}
\end{equation*}
For the upper bound of $u_{\textup{RSE}}(\delta)$, 
notice that $\BR(\bvec{x}) \subseteq \BR_{\delta}(\bvec{x})$ for any $\delta > 0, \bvec{x} \in \Delta^m$. It then follows that 
\begin{multline*}\label{eq:delta-vs-SSE-utility}
        u_{\textup{SSE}} = \max_{\bvec{x} \in \Delta^{m}} \max_{j \in \BR(\bvec{x})} u_l(\bvec{x}, j) 
        \geq \max_{\bvec{x} \in \Delta^{m}} \min_{j \in \BR(\bvec{x})} u_l(\bvec{x}, j)
        \\ 
        \geq \max_{\bvec{x} \in \Delta^{m}} \min_{j \in \BR_{\delta}(\bvec{x})} u_l(\bvec{x}, j) 
        = u_{\textup{RSE}}(\delta).
\end{multline*}

\smallskip

Moreover, $u_{\textup{RSE}}(0^+)$ exists because $u_{\textup{RSE}}(\delta)$ is monotone non-increasing in $\delta$ and is upper bounded by $u_{\textup{SSE}}$. $u_{\textup{RSE}}(0^+) = u_{\textup{SSE}}$ is implied by the squeeze theorem, based on the bounds from Equations  \eqref{eq:property1part1} and~\eqref{eq:property1part2} from Property~\ref{rse-property-2} when $\Delta > 0$. 
In Appendix \ref{sec:appendix-proof-section-property1}, we show an instance with $u_{\textup{RSE}}(0^+) < u_{\textup{SSE}}$ when $\Delta = 0$. 

\smallskip
\noindent{\bf Property \ref{rse-property-2}.}  
To prove Equation~\eqref{eq:property1part2}, we construct a leader strategy $\hat{\bvec{x}}$ and show that playing $\hat{\bvec{x}}$ against a $\delta$-rational follower yields a utility of at least $u_{\textup{SSE}} - \frac{\delta}{\Delta}$ for the leader.
Let $( \bvec{x}^*, j^*)$ be the SSE of the game. 
Let $\bvec{x}^{j^{*}}$ be a strategy such that $u_f(\bvec{x}^{j^{*}}, j^*) \geq u_f(\bvec{x}^{j^{*}}, j') + \Delta$ for any $j' \neq j^*$. By definition of the inducibility gap, such a strategy must exist. 

Since $\Delta > \delta$, we can set $\hat{\bvec{x}} = (1 - \frac{\delta}{\Delta}) \bvec{x}^* + \frac{\delta}{\Delta} \bvec{x}^{j^{*}}$, which is a valid leader strategy. We have $\BR_{\delta}(\hat{\bvec{x}}) = \{j^*\}$, since the following inequality holds for any $j' \neq j^*$,
\begin{equation*}\label{eq:prop_comb_strategy}
    \begin{split}
        u_f(\hat{\bvec{x}}, j^*) &= (1 - \frac{\delta}{\Delta}) u_f(\bvec{x}^*, j^*) +\frac{\delta}{\Delta}u_f(\bvec{x}^{j^{*}}, j^*) \\
        &\geq (1-\frac{\delta}{\Delta})u_f(\bvec{x}^*, j')  + \frac{\delta}{\Delta} \big(u_f(\bvec{x}^{j^{*}}, j') + \Delta\big)  \\
        &=  (1-\frac{\delta}{\Delta})u_f(\bvec{x}^*, j')  + \frac{\delta}{\Delta} u_f(\bvec{x}^{j^{*}}, j') + \delta \\
        &=u_f(\hat{\bvec{x}}, j') + \delta.
    \end{split}
\end{equation*}
Hence, we have $\min_{j \in \BR_{\delta}(\hat{\bvec{x}})} u_l(\hat{\bvec{x}}, j) = u_l(\hat{\bvec{x}}, j^*)$, which can  be bounded from below as 
\begin{multline*}\label{eq:propFinalStep}
        u_l(\hat{\bvec{x}}, j^*) = (1 - \frac{\delta}{\Delta}) u_l(\bvec{x}^*, j^*) +\frac{\delta}{\Delta}u_l(\bvec{x}^{j^{*}}, j^*)  
        \geq (1 - \frac{\delta}{\Delta}) u_l(\bvec{x}^*, j^*) 
        = (1 - \frac{\delta}{\Delta}) u_{\textup{SSE}},
\end{multline*}
where we used $u_l(\bvec{x}^{j^{*}}, j^*) \geq 0$. 
Since $u_{\textup{SSE}} \le 1$, we have 
\[
u_{\textup{RSE}}(\delta) 
\geq \min_{j \in \BR_{\delta}(\hat{\bvec{x}})} u_l(\hat{\bvec{x}}, j)
\geq (1 - \delta/\Delta) u_{\textup{SSE}} 
\geq u_{\textup{SSE}} - \delta/\Delta.
\]

\paragraph{\bf Property \ref{rse-property-3}.}
We demonstrate with examples in Appendix~\ref{sec:appendix-proof-section-property3} on how $u_{\textup{RSE}}(\delta)$ may be discontinuous when $\delta \geq \Delta$.
In what follows we prove its Lipschitz continuity for $\delta \in (0, \Delta)$.
  Pick arbitrary $L > 1/ \Delta$ and two arbitrary numbers $\delta$ and $\delta'$ such that $0 < \delta < \delta' \le \Delta - 1/L$. %
We show that 
$|u_{\textup{RSE}}(\delta) - u_{\textup{RSE}}(\delta')| \le L (\delta' - \delta)$ to complete the proof.

Let $(\bvec{x}^*, j^* )$ be a $\delta$-RSE.
Pick arbitrary 
$\tilde{j} \in \argmax_{j \in \BR_{\delta}(\bvec{x}^*)}  u_f(\bvec{x}^*, j)$,
and let $\tilde{\bvec{x}}$ be a strategy such that 
$u_f(\tilde{\bvec{x}}, \tilde{j}) \geq u_f(\tilde{\bvec{x}}, j) + \Delta$
for all $j \neq \tilde{j}$
(which exists according to the definition of the inducibility gap). 
We construct a leader strategy 
$
\hat{\bvec{x}} = \frac{\Delta - \delta'}{\Delta - \delta} \bvec{x}^* + \frac{\delta' - \delta}{\Delta - \delta} \tilde{\bvec{x}}.
$
We have $j \notin \BR_{\delta'}(\hat{\bvec{x}})$ 
if $j \notin \BR_\delta(\bvec{x}^*)$ because
\begin{equation}\label{eq:deltaPlusResponseSet}
    \begin{split}
        u_f(\hat{\bvec{x}}, \tilde{j}) 
        &= \frac{\Delta - \delta'}{\Delta - \delta} u_f(\bvec{x}^*, \tilde{j}) +\frac{\delta'-\delta}{\Delta - \delta} u_f(\tilde{\bvec{x}}, \tilde{j}) \\
        &\geq \frac{\Delta - \delta'}{\Delta - \delta}\left( u_f(\bvec{x}^*, j) + \delta\right) + \frac{\delta'-\delta}{\Delta - \delta} \left(u_f(\tilde{\bvec{x}}, j) + \Delta\right) \\
        &= \frac{\Delta - \delta'}{\Delta - \delta}u_f(\bvec{x}^*, j)  + \frac{\delta'-\delta}{\Delta - \delta} u_f(\tilde{\bvec{x}}, j) + \delta' \\
        &=u_f(\hat{\bvec{x}}, j) + \delta',
    \end{split}
\end{equation}
where $u_f(\bvec{x}^*, \tilde{j}) \ge u_f(\bvec{x}^*, j) + \delta$ because $\tilde{j} \in \argmax_{j \in \BR_{\delta}(\bvec{x}^*)}  u_f(\bvec{x}^*, j)$, $j \notin \BR_\delta(\bvec{x}^*)$.
Hence, $\BR_{\delta'}(\hat{\bvec{x}}) \subseteq \BR_\delta(\bvec{x}^*)$ and  
\[
\begin{split}
u_{\textup{RSE}}(\delta') 
= \min_{j \in \BR_{\delta'}(\hat{\bvec{x}})} u_l(\hat{\bvec{x}}, j) 
&= \min_{j \in \BR_{\delta'}(\hat{\bvec{x}})} \left( \frac{\Delta - \delta'}{\Delta - \delta}u_l(\bvec{x}^*, j)  + \frac{\delta'-\delta}{\Delta - \delta} u_l(\tilde{\bvec{x}}, j) \right) \\ 
&\geq \min_{j \in \BR_\delta(\bvec{x}^*)} \frac{\Delta - \delta'}{\Delta - \delta}u_l(\bvec{x}^*, j) 
= \frac{\Delta - \delta'}{\Delta - \delta} u_{\textup{RSE}}(\delta).
\end{split}
\]
This means that 
\[
u_{\textup{RSE}}(\delta) - u_{\textup{RSE}}(\delta') \le \frac{\delta' - \delta}{\Delta - \delta} \cdot u_{\textup{RSE}}(\delta) \le \frac{\delta' - \delta}{\Delta - \delta} \le L (\delta' - \delta),
\]
where the last inequality is due to $\delta \le \Delta - 1/ L$.
Since $u_{\textup{RSE}}$ is non-increasing as we argued for Property~(1), we then have $|u_{\textup{RSE}}(\delta) - u_{\textup{RSE}}(\delta')| = u_{\textup{RSE}}(\delta) - u_{\textup{RSE}}(\delta') \le L (\delta' - \delta)$.

\qed

We conclude this section with a discussion on the convexity of  $u_{\textup{RSE}}(\delta)$. Intuitively, one would conjecture that $u_{\textup{RSE}}(\delta)$ is convex non-increasing, as the leader's payoff should have diminishing margin as $\delta$ increases and follower chooses the worse responses.  
However, this intuition is only correct when $m=2$. The function in general is neither convex nor concave, as illustrated in the plot in Figure \ref{fig:rse}.

\begin{proposition}[(non-)convexity of $u_{\textup{RSE}}(\delta)$]\label{prop:rse-convexity}
When $m=2$, $u_{\textup{RSE}}(\delta)$ is convex when $\delta \in (0, \Delta)$. When $m>2$, there exist game instances where $u_{\textup{RSE}}(\delta)$ is neither convex nor concave, even when $\delta \in (0, \Delta)$.      
\end{proposition}

The proof hinges on the fact that, when $m=2$, $u_{\textup{RSE}}(\delta)$ is the pointwise maximum of a set of linear functions, each of which corresponds to the optimal leader utility in a $\delta$-optimal response region. We also construct an explicit example of neither convex nor concave $u_{\textup{RSE}}(\delta)$ with $m=3$. See Appendix~\ref{appendix:rse-convex} for the full proof of Proposition~\ref{prop:rse-convexity}.

\section{Computational Complexity of RSE}\label{sec:computational}%
In this section, we study the complexity of computing and approximating a $\delta$-RSE. 
We first show that NP-hardness of the computation problem in general and then propose a QPTAS algorithm to compute the approximated $\delta$-RSE.

\subsection{Hardness of Approximating $\delta$-RSE}

We start with the result that it is NP-hard, in general, to obtain an $\epsilon$-optimal $\delta$-RSE leader strategy.
We remark that, though $\delta$-RSE appears to be a  natural solution concept, we are not aware of any previous results on its computational complexity. 
The closest result we can find is the NP-hardness of computing an optimal leader strategy that is robust with respect to uncertainty about the follower's utility matrix, studied by \cite{letchford2009learning}.  
Despite some similarity in spirit, these two problems can not be seen as special cases of each other. 
Moreover, from a technical point of view, our hardness result also sheds light on the inapproximability of the problem whereas the proof technique of \cite{letchford2009learning} only implies the hardness of exact computation and leaves inapproximability an open problem from their problem.

\begin{theorem}\label{thm:hardness}
It is NP-hard to compute a $\frac{1}{2n}$-optimal $\delta$-RSE leader strategy.
\end{theorem}

\smallskip
\noindent\textit{Proof of Theorem \ref{thm:hardness}.}\quad
We show a reduction from the {\sc eXact 3-set Cover (X3C)} problem.
An {\sc X3C} instance is given by an integer $k$, a collection of $m$ subsets $S_1, \dots, S_m \subseteq [3k]$, each of size 3. 
It is a yes-instance if there exists $J \subseteq [m]$, such that $|J| = k$ and $\bigcup_{j \in J} S_j = [3k]$; we call such a $J$ an {\em exact cover}. Otherwise, it is a no-instance. 
We reduce an instance of {\sc X3C} to a game with the following utility matrices.
The leader has $m$ actions to choose from, each corresponding to a subset in the  {\sc X3C} instance.
The follower has $n = m+ 3k + 1$  actions $\{a\} \cup \{b_j: j \in [m]\}  \cup \{c_i: i \in [3k]\}$, where each $b_j$ corresponds to subset $S_j$, and each $c_i$ corresponds to an element in $[3k]$.

\begin{figure}[tbh]
\includegraphics[width=1\textwidth]{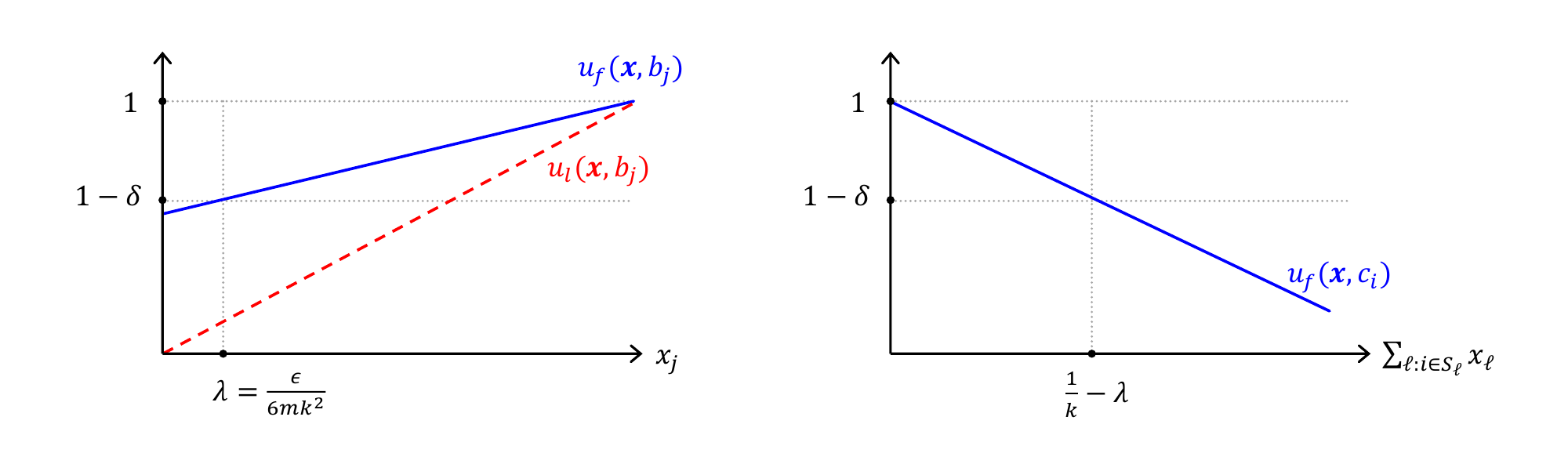}
\caption{Utility Functions of Constructed Instances for the Reduction of Theorem \ref{thm:hardness}.\label{fig:reduction}}
\end{figure}

Suppose $\epsilon > 0$ is a constant,
 and let $\lambda = \frac{\epsilon}{6m \cdot k^2}$. The follower's utility function is given as follows (also see Figure~\ref{fig:reduction} for an illustration).
\begin{itemize}
\item For all $\ell \in [m]$:
$u_f(S_\ell, a) = 1$.

\item For all $\ell \in [m]$ and $j \in [m]$:
\begin{align}
\label{eq:uf-bb}
u_f(S_\ell, b_j) =
\begin{cases}
\max \left\{ 1 -   \frac{\delta}{1 - \lambda}, \quad 0 \right\}, & \text{ if } j \neq \ell \\
\min \left\{1, \quad \frac{1-\delta}{\lambda} \right\}, & \text{ if } j = \ell
\end{cases}
\end{align}
This ensures that $u_f(\bvec{x}, b_j) \le 1 - \delta$ whenever $x_j \le \lambda$, and $1 - \delta < u_f(\bvec{x}, b_j) \le 1$, otherwise.

\item For all $\ell \in [m]$ and $i \in [3k]$:
\begin{align}
\label{eq:uf-c}
u_f(S_\ell, c_i) =
\begin{cases}
\min \left\{ (1-\delta) \cdot \frac{k}{k-1 +\lambda \cdot k}, \quad 1 \right\}, & \text{ if } i \notin S_\ell \\
\max \left\{0, \quad  1 - \delta \cdot \frac{k}{1 - \lambda \cdot k} \right\}, & \text{ if } i \in S_\ell 
\end{cases}
\end{align}
This ensures that $1 - \delta < u_f(\bvec{x}, c_i) \le 1$ whenever $\sum_{\ell: i \in S_\ell} x_\ell < \frac{1}{k} - \lambda$, and $u_f(\bvec{x}, c_i) \le 1 - \delta$ otherwise.
\end{itemize}

\noindent The leader's utility function is given as follows.
\begin{itemize}
\item 
For all $\ell \in [m]$:
\begin{align}
\label{eq:ul-ac}
u_l(S_\ell, a) = \frac{1}{k},   
\quad \text{ and } \quad 
u_l(S_\ell, c_i) = 0\quad  \text{ for all } i \in [3k] 
\end{align}

\item
For all $\ell \in [m]$ and $j \in [m]$:
\begin{align}
\label{eq:ul-bb}
u_l(S_\ell, b_j) =
\begin{cases}
0 & \text{ if } j \neq \ell \\
1 & \text{ if } j = \ell 
\end{cases}
\end{align}
\end{itemize}

We show that if the {\sc X3C} instance is a yes-instance, then the leader obtains utility $\frac{1}{k}$ in a robust Stackelberg equilibrium;
otherwise, the leader obtains at most $\frac{1}{2k} \cdot(1 + \epsilon)$.
Hence, no $\frac{1}{2k}\cdot(1- \epsilon)$-optimal algorithm exists unless P = NP. 
Intuitively, to obtain the higher utility of $\frac{1}{k}$, the leader needs to prevent the follower's actions $c_i$ from being a $\delta$-optimal response.
According to the utility definition, this requires choosing actions $S_\ell$ that cover $i$ with a sufficiently high probability close to $1/k$; see Figure~\ref{fig:reduction} (right). 
On the other hand, choosing each $S_\ell$ comes with a price as it will cause $b_\ell$ to be a $\delta$-optimal response if the probability reaches $\lambda$; see Figure~\ref{fig:reduction} (left).
Hence, in order to maintain utility $\frac{1}{k}$ for the leader, for each $S_\ell$, we should either pick it with probability close to zero (i.e., $< \lambda$), or we pick it with probability at least $\frac{1}{k}$. This is analogous to a discrete choice of $S_\ell$ as in the {\sc X3C}.
More specifically, the reduction proceeds as follows.

First, suppose that the {\sc X3C} instance is a yes-instance and $J$ is an exact cover of this instance.
Consider the following leader strategy $\bvec{x} = (x_j)_{j\in [m]}$, whereby the leader plays each pure strategy $S_j$ with probability: $x_j = \frac{1}{k}$ if $j \in J$, and $x_j = 0$ if $j \notin J$.
The follower's utility for responding to $\bvec{x}$ with each pure strategy is as follows:
\begin{itemize}
\item 
Clearly, $u_f(\bvec{x}, a) = 1$ and $u_l(\bvec{x}, a) = \frac{1}{k}$.
\item
For each $b_j$, according to \eqref{eq:uf-bb}:
\begin{itemize}
\item    
If $j \in J$, we have $x_j = \frac{1}{k} > \lambda$ and hence, 
$u_f(\bvec{x}, b_j) > 1 - \delta$; 
Meanwhile, $u_l(\bvec{x}, b_j) = \frac{1}{k}$ according to \eqref{eq:ul-bb}.
\item
If $j \notin J$, we have $x_j = 0 < \lambda$ and hence, 
$u_f(\bvec{x}, b_j) < 1 - \delta$.
\end{itemize}

\item 
For each $c_i$,  we have $i \in S_j$ for some $j \in J$ since $J$ is an exact cover. Hence, $\sum_{\ell: i \in S_\ell} x_\ell \ge \frac{1}{k} > \frac{1}{k} - \lambda$, and we have
$u_f(\bvec{x}, c_i) < 1 - \delta$.
\end{itemize}

\noindent As a result, $\BR_{\delta}(\bvec{x}) = \{ a\} \cup \{b_j : j \in J \}$, and $\min_{j' \in \BR_{\delta}(\bvec{x})} u_l(\bvec{x}, j') = \frac{1}{k}$.

\medskip

Conversely, suppose that by playing some strategy $\bvec{x}$, the leader obtains utility at least $\frac{1}{2k} \cdot(1 + \epsilon)$.
We show that the instance must be a yes-instance.
In this case, we have $\min_{y \in \BR_{\delta}(\bvec{x})} u_l(\bvec{x}, y) \ge \frac{1}{2k} \cdot(1 + \epsilon)$. According to the definition of the leader's utility function, this implies:
\begin{itemize}
\item
$c_i \notin \BR_{\delta}(\bvec{x})$ for all $i \in [3k]$.
Hence, we have 
\begin{align}
\label{eq:reduction-xj-lb}
\sum_{\ell: i \in S_\ell} x_\ell \geq  \frac{1}{k}- \lambda.
\end{align}
Since $|S_\ell| = 3$, we have 
$\sum_{i\in[3k]} \sum_{\ell: i \in S_\ell} x_\ell 
= 3 \sum_{\ell \in [m]} x_\ell 
= 3$, so it holds for all $i \in [3k]$ that:
\begin{align}
\label{eq:reduction-xj-ub}
\sum_{\ell: i \in S_\ell} x_\ell 
= 3 - \sum_{i'\in[3k]\setminus\{i\}} \sum_{\ell: i' \in S_\ell} x_\ell 
< \frac{1}{k} + 3k \cdot \lambda
\le \frac{1}{k} \cdot (1 + \epsilon/2).
\end{align}

\item
If $b_j \in \BR_{\delta}(\bvec{x})$, then it must be that $x_j \ge \frac{1}{2k} \cdot(1 + \epsilon)$.
Hence, for each $j \in [m]$, either $x_j \ge \frac{1}{2k} \cdot(1 + \epsilon)$, or $x_j \le \lambda$ (as implied by $b_j \notin \BR_{\delta}(\bvec{x})$).
This further implies that for each $i \in [3k]$, there is exactly one $\ell$ with $i \in S_\ell$ and $x_\ell \ge \frac{1}{2k} \cdot(1 + \epsilon)$:
the existence of two or more such $\ell$ would violate \eqref{eq:reduction-xj-ub};
on the other hand, if no such $\ell$ exists, 
we would have $\sum_{\ell: i \in S_\ell}x_\ell \leq m \cdot \lambda < \frac{1}{k} - \lambda$, which violates \eqref{eq:reduction-xj-lb}.
It follows that, for this $\ell$, we have 
\begin{align*}
x_\ell 
&= 
\sum_{\ell' : i \in S_{\ell'}} x_{\ell'} - \sum_{\ell': \ell' \neq \ell \text{ and } i \in S_{\ell'}} x_{\ell'} \\
&\ge 
\sum_{\ell' : i \in S_{\ell'}} x_{\ell'} - (m-1) \cdot \lambda
\quad \ge \quad 
1/k - m \cdot \lambda 
\quad > \quad 
1/k - \epsilon/ k^2,
\end{align*}
which implies that the set $J := \left\{ \ell \in [m]: x_\ell \ge \frac{1}{2k} \cdot(1 + \epsilon) \right\}$ has at most $k$ element in it:
otherwise, $\sum_{\ell \in J} x_\ell > (k+1) \cdot (1/k - \epsilon / k^2) > 1$. 
Therefore, $J$ is an exact cover, and the {\sc X3C} instance is a yes-instance.
\end{itemize}
\noindent Therefore, we have shown that no $\frac{1}{2k}\cdot(1- \epsilon)$-optimal algorithm exists unless P = NP. Since $n>3k$, it is also NP-hard to compute $\frac{1-\epsilon}{n}$-optimal $\delta$-RSE leader strategy for any $\epsilon \in (0, 1]$. 
This completes the proof.
\qed

Theorem \ref{thm:hardness} shows that it is NP-hard in general to approximate a $\delta$-RSE. 
However, as a corollary of Property (1) in Theorem \ref{thm:delta-RSE-property},
we show that there exists an efficient algorithm to compute  $\frac{\delta}{\Delta}$-optimal $\delta$-RSE leader strategy. This intuitively illustrates that  the difficult instances of finding $\delta$-RSE are those with small inducibility gap $\Delta$ but large robustness requirement $\delta$.  We refer  readers to the detailed algorithm of this corollary in Appendix \ref{appendix-computational-obv}

\rev{
 \begin{corollary}
 \label{obv:efficent-rse}
For Stackelberg games with inducibility gap $\Delta > \delta$, there exists an algorithm that computes $\frac{\delta}{\Delta}$-optimal  $\delta$-RSE leader strategy in   $O(\poly(m,n))$ time.
\end{corollary} 
The suboptimality gap $\delta/\Delta$ in Corollary~\ref{obv:efficent-rse} can be arbitrarily large when $\Delta$ is close to $0$, in which case the approximation would not be useful. Hence, we will next aim to develop an algorithm whose approximation ratio is irrespective of the inducibility gap~$\Delta$.
}

\subsection{A QPTAS for $\delta$-RSE} 

We now present a quasi-polynomial time approximation scheme (QPTAS) for computing an approximate $\delta$-RSE, for any given $\delta$. 
\rev{
The approximation ratio $\delta$ holds irrespective of any specific properties of the game, such as Lipschitz continuity or the inducibility gap, so the algorithm is applicable to a broad range of instances.
}
We also leave an intriguing open problem here to close the gap between the efficiency of this algorithm and the above inapproximability result --- specifically,  to understand whether a PTAS exists for $\delta$-RSE or whether  Theorem \ref{thm:hardness} can be strengthened to the hardness of obtaining a constant additive approximation (possibly under some assumption like  exponential time hypothesis as used by \citet{rubinstein2016settling} to rule out PTAS for Nash equilibrium). Our preliminary investigation suggests that either direction seems to require significantly different ideas from our current techniques.

\begin{theorem}
\label{thm:epsilon-approx-utility}
For any $\epsilon > 0$, we can compute an $\epsilon$-optimal  $\delta$-RSE leader strategy in quasi-polynomial time $O\big( m^{\left\lceil \frac{\log 2n}{2\epsilon^2}\right\rceil} n \log n  \big)$. 
\end{theorem}

Before presenting the formal proof, we briefly overview the high-level idea. Our algorithm starts with a probabilistic argument similar to \cite{lipton2003playing} for arguing the existence of an approximate Nash equilibrium with a simple format termed the $k$\emph{-uniform strategy}. Formally, a mixed strategy $\bvec{x} \in \Delta^{m}$ is called $k$-uniform for some integer $k$ if every $x_i = k_i/k$ for some integer $k_i \leq k$. \citet{lipton2003playing} prove that in any two-player $m\times m$ matrix game there always exists a pair of $k$-uniform strategies for $k = \frac{12 \ln m}{\epsilon^2}$ that is an $\epsilon$-Nash equilibrium. 
Consequently, to find an $\epsilon$-Nash, one only needs to exhaustively search all  possible $k$ uniform mixed strategy pairs. Searching for an $\epsilon$-optimal $\delta$-RSE, however, requires significantly more work due to the bi-level nature of our problem. 
\rev{
In fact, $\epsilon$-optimal $\delta$-RSEs may not be $k$-uniform strategies in general:
a $k$-uniform leader strategy $\bar{\bvec{x}}$ may induce a very different $\delta$-best response set, and consequently a different worst-case follower response, from the ones induced by a leader strategy nearby, which makes the leader's utility discontinuous over the strategy space.
}
Consequently, the main part of our proof is to efficiently search, through a carefully crafted binary search procedure,  the entire \emph{nearby convex region} of  each uniform strategy $\bar{ \bvec{x} }$, in order to identify a leader strategy $\bvec{x}$ that induces the most favorable follower response action. We present the details in the proof below. 

\smallskip
\noindent\textit{Proof of Theorem \ref{thm:epsilon-approx-utility}.}\quad
Let $\mathcal{G}_k \subseteq \Delta^m$ denote the set of all $k$-uniform mixed strategies for the leader (who has $m$ actions). Note that there are $O(m^k)$ many $k$-uniform strategies\footnote{The total number can be computed as dividing $k$ items into $m$ parts, while each part can have zero item.}. 
The following lemma is originally from \citet{althofer1994sparse} and later used by \citet{lipton2003playing} for computing approximate Nash equilibrium. It can be proved via the probabilistic method \cite{alon2016probabilistic}. 

\begin{lemma}[\citet{althofer1994sparse,lipton2003playing}]\label{lem:approx_strategy_space_main}
Let $A \subseteq [0,1]^{m\times n}$ be the leader's   payoff matrix. For any $\epsilon>0$ and any leader strategy $\bvec{x} \in \Delta^{m}$, there exists a $k$-uniform strategy $\bar{\bvec{x}}  \in \mathcal{G}_k $ with $k=\lceil \frac{\log \, 2n}{2\epsilon^2}\rceil$    such that \[|u_l(\bvec{x},j) - u_l(\bar{\bvec{x}},j)| \leq \epsilon\quad \text{  for all } j=1,\cdots,n. \]
\end{lemma} 

We now use Lemma $\ref{lem:approx_strategy_space_main}$ to construct subspaces of the leader's strategy space. Specifically, for each $\bar{\bvec{x}} \in \mathcal{G}_k$, we construct $\Delta^{\bar{\bvec{x}}} \subseteq \Delta^{m}$ such that: \[\Delta^{\bar{\bvec{x}}} = \big\{\bvec{x} \big| \bvec{x} \in \Delta^{m} \text{ and } |u_l(\bvec{x},j) - u_l(\bar{\bvec{x}},j)| \leq \epsilon \text{  for all } j=1,\cdots,n\big\}\]

Note that each $\Delta^{\bar{\bvec{x}}}$ is a convex region since it is defined by a set of linear constraints by writing $|u_l(\bvec{x},j) - u_l(\bar{\bvec{x}},j)| \leq \epsilon$ as $u_l(\bvec{x},j) - u_l(\bar{\bvec{x}},j)  \leq \epsilon$ and $-u_l(\bvec{x},j) + u_l(\bar{\bvec{x}},j)  \leq \epsilon $.   Moreover, $\bigcup_{\bar{\bvec{x}} \in \mathcal{G}_k} \Delta^{\bar{\bvec{x}}} = \Delta^{m}$, because  Lemma \ref{lem:approx_strategy_space_main} implies that any $  \bvec{x} \in \Delta^{m}$ belongs to some $\Delta^{\bar{\bvec{x}}}$.

The key to our proof is to compute an approximately optimal $\delta$-RSE leader strategy within each convex region    $\Delta^{\bar{\bvec{x}}}$. Note that, fixing any follower response action $j$,  the mixed strategies $\bvec{x} \in \Delta^{\bar{\bvec{x}}}$ gives rise to roughly the same leader utility, up to at most $\epsilon$ difference by Lemma \ref{lem:approx_strategy_space_main}. However, this does not imply that they are   equally good since different mixed strategies may lead to different sets $\BR_{\delta}(\bvec{x})$ of $\delta$-optimal follower responses, which in turn induces   different leader utilities. So we need to search for the $\bvec{x} \in \Delta^{\bar{\bvec{x}}}$ to maximize the worst (over possible follower responses) leader utility, or formally to solve the following \begin{equation}\label{eq:binary-search-prob}
 \text{optimization problem within }\Delta^{\bar{\bvec{x}}}: \qquad   \max_{ \bvec{x} \in \Delta^{\bar{\bvec{x}}} }  \min_{j \in \BR_{\delta}(\bvec{x})} u_l(\bvec{x}, j) 
\end{equation} 

Unfortunately, Problem \eqref{eq:binary-search-prob} is generally intractable since the feasible region of inside  $\min$ depends on $\bvec{x}$. We instead solve the following more tractable variant 
\begin{equation}\label{eq:binary-search-prob2}
\qquad  \text{surrogate of Problem \eqref{eq:binary-search-prob}}: \qquad   \max_{ \bvec{x} \in \Delta^{\bar{\bvec{x}}} }  \min_{j \in \BR_{\delta}(\bvec{x})} u_l(\bar{\bvec{x}}, j) 
\end{equation} 
which substitutes $u_l(\bvec{x}, j)$ in  Problem \eqref{eq:binary-search-prob} by $u_l(\bar{\bvec{x}}, j)$. Observe that any optimal solution to Problem \eqref{eq:binary-search-prob2} must be an $\epsilon$-optimal solution to Problem \eqref{eq:binary-search-prob} because their objective function differs by at most $\epsilon$ due to  Lemma \ref{lem:approx_strategy_space_main} and our restriction of $\bvec{x} \in \Delta^{\bar{\bvec{x}}}$.   

What is nice about Problem \eqref{eq:binary-search-prob2} is that its objective function  only directly depends on $j$ (whose choice then depends on $\bvec{x}$). This allows us to design the Algorithm \ref{alg:exist-utility} that can efficiently search for the best $\bvec{x} \in \Delta^{\bar{\bvec{x}}}$. We prove its correctness in the following lemma.

\begin{algorithm}[tbh]
    \caption{\textsc{Utility-Verification}}
    \label{alg:exist-utility}

\SetKwInOut{Input}{Input}
\SetKwInOut{Output}{Output}

\Input{Leader strategy $\bar{\bvec{x}}\in \mathcal{G}_k$,   its corresponding  $\Delta^{\bar{\bvec{x}}}$, and  target utility $\mu$.}

\Output{If $\exists \, \bvec{x} \in \Delta^{\bar{\bvec{x}}}$ such that $\min_{j \in \BR_{\delta}(\bvec{x})} u_l(\bar{\bvec{x}}, j) \geq \mu$, output \texttt{True}  and such an $\bvec{x}$; else output \texttt{False}.}

\BlankLine

$Q \leftarrow \varnothing$;

\For{every follower action $j \in [n]$}
{
    \If{$u_l(\bar{\bvec{x}},j)   < \mu$}
    {
    $Q \leftarrow Q \union \{j\}$
    }
}

\For{every follower action $j \in [n]$ and $j \notin Q$}
{
Determine if the following Linear Program is feasible:
    \begin{equation}\label{LP:exist-search}
        \begin{split}
            \exists \quad &\bvec{x} \in \Delta^{\bar{\bvec{x}}} \\
            \text{s.t.} \quad &u_f(\bvec{x}, j) \geq u_f(\bvec{x}, j'), \forall j'\in[n] \\
            &u_f(\bvec{x}, j) \geq u_f(\bvec{x}, j') + \delta, \forall j'\in Q
        \end{split}
    \end{equation}
\If{the above linear feasibility problem is feasible for some $j$}
{\Return \texttt{True} and any feasible solution $\bvec{x} $ of that problem.}
}

\Return \texttt{False}.
\end{algorithm}

\begin{lemma}\label{lem:exist-search}
For any $\mu \in [0,1]$, there is a polynomial time algorithm that asserts whether the optimal objective of Problem \eqref{eq:binary-search-prob2} is larger than $\mu$ or not, and in the former case outputs an $\bvec{x} \in \Delta^{\bar{\bvec{x}}}$ that achieves  $\min_{j \in \BR_{\delta}(\bvec{x})} u_l(\bar{\bvec{x}}, j) \geq \mu$. 
\end{lemma}
\noindent\textit{Proof.}
The details of this algorithm are presented in Algorithm \ref{alg:exist-utility}. At a high level, we first identify all ``bad'' follower actions $j$'s that cannot satisfy our request, i.e., $u_l(\bar{\bvec{x}},j)   < \mu$, and group them into set $Q$. We then try to see whether there exists an  $\bvec{x} \in \Delta^{\bar{\bvec{x}}}$ such that its follower $\delta$-optimal response set does not contain any bad actions, i.e., $\BR_{\delta}(\bvec{x}) \cap Q = \varnothing$. This later question reduces to a series of linear feasibility problems, each for a $j \not \in  Q$ (the Program \eqref{LP:exist-search}) deciding   whether there exists a $\bvec{x} \in \Delta^{\bar{\bvec{x}}}$ under which the $j$ is the best follower action and the follower's utility from $j$ is at least $\delta$ larger than his utility from any $j' \in Q$.  
\qed

Armed with Lemma \ref{lem:exist-search}, we can use binary search to find an $\bvec{x}$ that exactly solves Problem~\eqref{eq:binary-search-prob2}  after $\log(1/n)$ rounds since we know the problem only has  $n$ possible values of $\mu$ (i.e. $u_l(\bar{\bvec{x}}, 1), \cdots, u_l(\bar{\bvec{x}}, n)$).   This solution will be  $\epsilon$-optimal for Problem \eqref{eq:binary-search-prob}. We do this for the $O(m^k)$ possible $k$-uniform strategies and output the strategy with the largest objective. This is  a   $\epsilon$-optimal $\delta$-RSE. 
\qed

\section{Statistical Complexity of RSE}\label{sec:learnability}%

In this section, we turn to another complexity study of RSE, i.e., the sample efficiency of learning an $\epsilon$-optimal $\delta$-RSE without initially knowing the leader's or follower's utility matrix.  Motivated by the recent work of learning the strong Stackelberg equilibrium (SSE)  by \citet{bai2021sample}, here we extend it to an online learning problem of the $\delta$-RSE. Similar to \cite{bai2021sample}, the learner cannot directly observe the mean reward matrix of a Stackelberg game $u_l, u_f \in \mathbb{R}^{m \times n}$, but has to learn to approximate the $\delta$-RSE from the noisy bandit feedback. 

\rev{The motivation for this learning paradigm is inspired by a common multi-agent learning practice of ``centralized training, decentralized execution''~\citep{lowe2017multi}, used in settings like robotics and game-playing (e.g., OpenAI Gym~\cite{brockman2016openai}). During training, a centralized algorithm can control the agents’ actions and access their feedback from the environment to learn game parameters and optimize their strategies. Once trained, the agents will deploy these strategies and operate independently in decentralized environments. In addition, this learning setup may capture scenarios where an environment simulator is available but sampling game outcomes of agents' strategy profile is costly --- for instance, in high-stakes strategic interactions or physical systems where each simulation run consumes significant resources.
We also remark that the notion of $\delta$-RSE is particularly natural in these real-world deployment, as followers may not play the exact best responses due to bounded rationality, noisy observations, or limited optimization capacity. Modeling such responses using the $\delta$-optimality assumption provides a more realistic and robust solution concept.
}

We describe the learning setting in Definition \ref{def:learning} and start this section by presenting a sample-efficient \rev{(yet not computationally efficient in general)} learning algorithm that can learn an approximate $\delta$-RSE with a utility guarantee. Notably, as a corollary, our sample complexity result strictly strengthens that of \cite{bai2021sample}, with both improved \emph{utility guarantee} and better \emph{computational  efficiency} for non-degenerate Stackelberg games (i.e., $\Delta > 0$)~\footnote{Interestingly, the algorithm for learning the mixed strategy SSE in  \cite{bai2021sample} happens to be solving an approximate $\delta$-RSE. Quoting the authors' own words in their paper, ``\emph{it is unclear whether this program (the $\delta$-RSE problem) can be reformulated to be solved efficiently in polynomial time}''. Our Theorem \ref{thm:hardness}  confirms that their program indeed is NP-hard.}. 

\begin{definition}[Learning $\delta$-RSE from Bandit Feedback \citep{bai2021sample}]
\label{def:learning}
 At each round, the learner can query an action pair $(i,j)$ and observe noisy bandit feedback, $r_l(i,j) = u_l(i,j) + \xi, r_f(i,j) = u_f(i,j) + \xi'$, where $\xi, \xi'$ are i.i.d. zero-mean noises with finite variances. 
\end{definition}

\begin{theorem}\label{thm:learning-rse}
There exists a learning algorithm that can learn an approximated $\delta$-RSE of any Stackelberg game $(u_l, u_f) \in \mathbb{R}^{m\times n}$  with leader's utility at least as much as $u_{\textup{RSE}}(\delta + 4\epsilon) - 2\epsilon$ using $O(mn\log(mn/\iota)/\epsilon^{2})$ samples with probability at least $1-\iota$. 
\end{theorem}
\noindent\textit{Proof of Theorem~\ref{thm:learning-rse}.}\quad
We prove its existence by explicitly constructing the learning algorithm. It starts with the following sampling procedure: 

Play each action pair $(i,j)$ for $T = \frac{1}{2\epsilon^2} \log(\frac{2mn}{\iota})$ rounds to get the mean reward estimation $ \tilde{u}_l(i,j) = \frac{1}{T}\sum_{t=1}^{T} r_l^t(i,j), \tilde{u}_f(i,j) = \frac{1}{T}\sum_{t=1}^{T} r_f^t(i,j)$.
According to the concentration inequality, both the estimation $\tilde{u}_l(i,j), \tilde{u}_f(i,j)$ have the error bound of $\epsilon$ with probability $1-\frac{\iota}{mn}$. Thus, by union bound, with probability $1- \iota$, the utility estimation $\tilde{u}_l, \tilde{u}_f$ satisfies $\norm{\tilde{u}_l - u_l}_{\infty} \leq \epsilon, \norm{\tilde{u}_f - u_f}_{\infty} \leq \epsilon$. The sample complexity in total is $\frac{mn}{2\epsilon^2} \log(\frac{2mn}{\iota}) = O(mn \log(mn/\iota)/\epsilon^{2} )$.

Notice that, for some leader strategy $\bvec{x}$, the $\delta$-best response set under $u_f$ can be different from that under $\tilde{u}_f$, which could result in \rev{a constant utility gap}. Hence, we set the benchmark to $(\delta + 2\epsilon)$-RSE in Lemma~\ref{lm:delta-eps-trade-off}.
\rev{Denote by $V(\bvec{x} ;u_l, u_f, \delta)$ the leader utility of strategy $\bvec{x}$ against a $\delta$-rational follower (who plays actions in the \rev{$\delta$-optimal} response set) in the game $({u}_l, {u}_f)$. Let $V^*(u_l, u_f, \delta)$ be a $\delta$-RSE of the game $({u}_l, {u}_f)$. }

\begin{lemma} \label{lm:delta-eps-trade-off}
For any $\norm{\tilde{u}_f - u_f}_{\infty} \leq \epsilon$, for any $\bvec{x} \in \Delta^m$, $V(\bvec{x} ; u_l, u_f, \delta)  
    \geq V(\bvec{x} ; u_l, \tilde{u}_f, \delta + 2\epsilon) $ and thus, $V^*( u_l, u_f, \delta)  
    \geq V^*(u_l, \tilde{u}_f, \delta + 2\epsilon) $.
\end{lemma}

\rev{We now construct a leader strategy based on the learnt Stackelberg game $(\tilde{u}_l, \tilde{u}_f)$ and bound its suboptimality with respect to a RSE of the true game $({u}_l, {u}_f)$.}
Let $(\bvec{x}, j^*)$ be the $(\delta+2\epsilon)$-RSE of the Stackelberg game $\tilde{u}_l, \tilde{u}_f$. We claim the learning algorithm could simply output $(\bvec{x}, j^*)$ as the approximated $\delta$-RSE of Stackelberg game $(u_l, u_f)$. 
The remaining proof is to prove the following series of inequalities:
\begin{align*}
    &V(\bvec{x} ; u_l, u_f, \delta)  
    \geq  V(\bvec{x} ; u_l, \tilde{u}_f, \delta + 2\epsilon) 
    \,\geq  V(\bvec{x} ; \tilde{u}_l, \tilde{u}_f, \delta + 2\epsilon) - \epsilon \\
    = \,  &    V^*(\tilde{u}_l, \tilde{u}_f, \delta + 2\epsilon) - \epsilon 
     \geq  V^*(\tilde{u}_l, u_f, \delta + 4\epsilon) - \epsilon 
     \geq  V^*(u_l, u_f, \delta + 4\epsilon) - 2\epsilon.
\end{align*}
There are four inequalities in the above arguments. The first and third inequality is by Lemma~\ref{lm:delta-eps-trade-off}. 
The second and last inequality is based on the fact that, for any $\norm{\tilde{u} - u}_{\infty} \leq \epsilon$, for any $\bvec{x} \in \Delta^m$, $\tilde{u}(\bvec{x},j) - u(\bvec{x},j) = \bvec{x}(\tilde{u} - u)e_j \in [-\epsilon, \epsilon]$.
The equality is by the construction of $\bvec{x}$ \rev{--- recall that $(\bvec{x}, j^*)$ is a $(\delta+2\epsilon)$-RSE of the Stackelberg game $(\tilde{u}_l, \tilde{u}_f)$.}
This concludes our main proof. We defer the proof of Lemma~\ref{lm:delta-eps-trade-off} to Appendix~\ref{sec:proof-delta-eps-trade-off}.

\qed

We make a few remarks on Theorem \ref{thm:learning-rse}.  First, we note that the above learning algorithm is sample-efficient but not computationally efficient, since it requires solving the exact $(\delta+2\epsilon)$-RSE of a Stackelberg game, which we already know is NP-hard to compute in general. However, it is possible to employ the QPTAS Algorithm~\ref{alg:exist-utility} to find an $\epsilon$-optimal $\delta$-RSE according to Theorem~\ref{thm:epsilon-approx-utility}, and this would not change the order of our learning algorithm's approximation ratio.\footnote{\rev{Specifically,
with Algorithm~\ref{alg:exist-utility}, we can find an $\epsilon$-optimal
$(\delta+2\epsilon)$-RSE of the Stackelberg game $(\tilde{u}_l, \tilde{u}_f)$ in quasi-polynomial time and this strategy would achieve a leader utility at least as much as $u_{\textup{RSE}}(\delta + 4\epsilon) - 3\epsilon$.
}
}

Second, Theorem~\ref{thm:learning-rse} combined with Theorem \ref{thm:delta-RSE-property} implies the following corollary about the efficient learning of an approximated SSE. 
Specifically, leveraging the convergence property of $\delta$-RSE, the following Corollary~\ref{col:learn-SSE} strengthens the SSE learning results in the recent work by \citet{bai2021sample}, in terms of providing better utility guarantee and computationally efficient learning algorithms. Specifically, \cite{bai2021sample} states that under the same sample complexity, a learning algorithm can only learn the  $u_{\textup{SSE}}$ up to $u_{\textup{RSE}}(\delta)$ while the gap between them can be arbitrarily large. 
But our result suggests, as long as the game instance is not degenerated (e.g., with inducibility $\Delta>0$), the  gap can be bounded so that SSE can be efficiently learned up to $\epsilon$ utility loss. 
Moreover, as highlighted by \cite{bai2021sample}, their learning algorithm  does not have a computational efficiency guarantee and may take exponential time in general, since there is no efficient algorithm to compute the approximated SSE with pessimistic follower tie-breaking. But our result implies that, as long as the game instance is non-degenerate with inducibility $\Delta>\epsilon$, we can efficiently compute $\epsilon$-RSE according to Corollary~\ref{obv:efficent-rse}.

\begin{corollary}[Efficient Learning of SSE] \label{col:learn-SSE}
For any Stackelberg game with inducibility $\Delta>0$, an $\epsilon$-optimal SSE can be learned from $O(\frac{1}{\epsilon^2})$ samples in polynomial time for any $\epsilon < \Delta$. %
\end{corollary}

\rev{ 
Thirdly, we explain how this learning result is almost tight due to the  Proposition \ref{prop:learningwithgap} below, where we present hard instances that fundamentally limit the learnability of $\delta$-RSE. 
In particular, the sample complexity dichotomy conditioned on whether $\Delta < \delta$ can be linked to the gap between a $\delta$-RSE and a $(\delta+4\epsilon)$-RSE  --- recall that the leader's utility guarantee in Theorem~\ref{thm:learning-rse} is given with respect to that of a $(\delta+4\epsilon)$-RSE.
In the case where $\Delta < \delta$, the $u_{\textup{RSE}}(\delta + 4 \epsilon)$  can be arbitrarily worse than $u_{\textup{RSE}}(\delta)$ due to the discontinuity of the $u_{\textup{RSE}}(\delta)$ function. 
This is intrinsic barrier can be further explained by the $\Omega(1)$ utility gap below. 
Meanwhile, in the case where $\Delta > \delta$, Property (3) of Theorem~\ref{thm:delta-RSE-property} allows us to use the Lipschitz continuity of $u_{\textup{RSE}}(\delta)$ to bound the gap between $\delta$-RSE and $(\delta+4\epsilon)$-RSE by $\epsilon$.
Specifically, given that $\Delta \geq \delta + 1/L$ for some constant $L$, the Lipschitz continuity of $u_{\textup{RSE}}(\delta)$ implies that $u_{\textup{RSE}}(\delta + 4 \epsilon)$ can be substituted by $u_{\textup{RSE}}(\delta) - O(\epsilon L)$ so that the sample complexity dependence on $\epsilon$ matches with the $\Omega(1/\sqrt{T})$ of the lower bound instance. 
We formalize the lower bound characterization in Proposition \ref{prop:learningwithgap} and its proof can be found in Appendix~\ref{appendix-learning-prop}.
}

\begin{proposition}\label{prop:learningwithgap}
For any sample size $T$, there exists a Stackelberg game instance with inducibility gap $\Delta$ such that any algorithm with $T$ samples in learning $\delta$-RSE is  $\Omega( 1/\sqrt{T} )$ suboptimal if $\delta < \Delta$, or $\Omega( 1 )$ suboptimal if $\delta \geq \Delta$.
More specifically, any output leader strategy $\hat{\bvec{x}}$ satisfies the following with probability at least $\frac{1}{3}$:
\begin{equation}\label{eq:observation2}
     \min_{j \in \BR_{\delta}(\hat{\bvec{x}})} u_l(\hat{\bvec{x}}, j) \leq 
     \begin{cases} 
     u_{\textup{RSE}}(\delta) - \frac{1/\sqrt{T} }{\Delta - \delta + 1/\sqrt{T} } & \delta < \Delta \\
     u_{\textup{RSE}}(\delta) - 1/2 & \delta \geq \Delta 
     \end{cases}.
\end{equation}
\end{proposition}

\section{Conclusion}
In practice, there are many situations where a follower fails to make the optimal decision in the Stackelberg game. However, the classic solution concept of SSE is not robust for suboptimal follower responses and leads to poor performance of the leader. In this paper, we systematically study a robust variant of Stackelberg equilibrium to account for suboptimal responses from the follower. We propose a well-defined definition of robust Stackelberg equilibrium, $\delta$-RSE, and show some nice properties of the leader's utility under a $\delta$-RSE. We identify the computational complexity for computing or approximating the $\delta$-RSE. We  show there does not exist an efficient algorithm for finding an approximate  $\delta$-RSE, unless P=NP, while we also propose a QPTAS to compute the $\epsilon$-optimal $\delta$-RSE for any $\epsilon > 0$. Lastly, we provide sample complexity results for learning the $\delta$-RSE when the follower utility is not known initially. 

Our results open up possibilities for many other interesting questions. For example, our positive and negative computational results in Section \ref{sec:computational} have a small gap due to the logarithmic exponent term in the computational complexity of the QPTAS. An immediate direction for future research is to close this gap by either showing a PTAS algorithm or strengthening the hardness result of inapproximability to a constant factor. In addition, our study concerns the most basic normal-form game setups. It is worth understanding the applicability of this robust solution concept to various applications of leader-follower or principal-agent game models, including pricing games~\citep{myerson1981optimal, devanur2014perfect}, security games~\citep{tambe2011security}, Bayesian persuasion~\citep{kamenica2011bayesian} and contract design~\citep{grossman1992analysis}. Future work can also study analogous robust solution concepts for more general Stackelberg game models such as these with Bayesian follower types~\citep{conitzer2006computing} or with constrained follower strategy sets~\citep{wald1945statistical,goktas2021convex}.  Lastly, we would also like to study the learnability of RSEs in different feedback models, e.g., when the learner cannot observe the follower payoffs \citep{blum2014learning, balcan2014learning, letchford2009learning, peng2019learning}.

\bibliographystyle{spbasic}      %
\bibliography{main}   %

\newpage

\appendix

\section{Omitted Proofs in Section \ref{sec:RSE-Property}}\label{appendix-property-section}

\subsection{Proof of Proposition \ref{prop:equilibrium_variants}}\label{appendix-property-suboptimal}
\noindent\textit{Proof.}\quad
Consider the Stackelberg game instance in Table \ref{table:Stackelberg_variants}, where both the leader and follower have 3 actions. In this game, we can compute the SSE leader strategy and max-min leader strategy as follows:
\begin{equation*}
\begin{split}
    {\bvec{x}}_1 = (1, 0, 0) = \argmax_{\bvec{x} \in \Delta^{m}} \max_{j \in \BR(\bvec{x})} u_l(\bvec{x}, j)  \quad \text{and} \quad
    {\bvec{x}}_2 = (0, 0, 1) = \argmax_{\bvec{x} \in \Delta^{m}} \min_{j \in [n]}  u_l(\bvec{x}, j) 
\end{split}
\end{equation*}
Meanwhile, the $\delta$-RSE leader strategy is $\bvec{x}^*=(0,1,0)$ and leader utility is $u_{\textup{RSE}}(\delta) = 2c$.
Next, we apply the SSE leader strategy and max-min leader strategy against the bounded rational follower. 
\begin{equation*}\label{eq:SE_variants}
    \begin{split}
        \min_{j \in \BR_{\delta}({\bvec{x}}_1)} u_l(\
    {\bvec{x}}_1, j) = u_l(i_1,j_2) = c \quad \text{and} \quad
        \min_{j \in \BR_{\delta}({\bvec{x}}_2)} u_l({\bvec{x}}_2, j) = u_l(i_3,j_1) = c
    \end{split}
\end{equation*}
while the $\delta$-RSE leader utility is $u_{\textup{RSE}}(\delta) = 2c$, causing a constant gap of $c$ between the $\delta$-RSE leader utility and the leader utility of playing SSE strategy or max-min strategy under the setting that allows approximately optimal follower responses. As we can set $c$ to be arbitrarily close to $1/2$, we have 
$$
\min_{j \in \BR_{\delta}({\bvec{x}}_1)} u_l({\bvec{x}}_1, j) < u_{\textup{RSE}}(\delta) - \frac{1}{2} \quad \text{\textit{and}} \quad
\min_{j \in \BR_{\delta}({\bvec{x}}_2)} u_l({\bvec{x}}_2, j) < u_{\textup{RSE}}(\delta) - \frac{1}{2}.
$$

\begin{table}[h]
\begin{center}
\begin{tabular}{ |c|c|c|c| } 
 \hline
  $u_l,u_f$& $j_1$ & $j_2$ & $j_3$ \\
 \hline
 $i_1$ &$1, \delta$ & $c, \delta$ & $0,0$\\ 
 \hline
 $i_2$ & $2c, \delta$ & $2c, \delta$ & $0,0$ \\
 \hline
 $i_3$ & $c, \delta$ & $c,\delta$ & $c,\delta$\\ 
 \hline
\end{tabular}
\end{center}
\caption{A Stackelberg game instance whose SSE leader strategy is $i_1$, $\delta$-RSE leader strategy is $i_2$ (for any $\delta > 0$), and max-min leader strategy is $i_3$, for any $c\in (0, 1/2)$.}
\label{table:Stackelberg_variants}
\end{table}

\qed

\subsection{Proof of Proposition~\ref{prop:tie-breaking}} \label{append:prop-tie-breaking}
\noindent\textit{Proof.}\quad
Consider a game with $m=3, n=2$ and utility matrices specified in Table~\ref{table:nonunique-delta-br-set}. Notice that the leader utility is maximized at the leader/follower action pair $(i_2, j_1)$, while the leader must place probability mass at least $\frac{\delta}{\Delta}$ on $i_1$, resulting in the leader utility no more than $\Delta - \delta$. In contrast, if the follower plays the action $j_2$, the leader can achieve a strictly better utility $\Delta - c$ given that $c < \delta$. Hence, $\forall \delta \in (0, \Delta)$, the $\delta$-RSE $(\bvec{x}^*, j^*)$ have leader strategy $ \bvec{x}^* = (0, 1, 0)$ and follower response $j^*=j_2$. We can compute for $\bvec{x}^*$ the leader and follower utility under different follower action, $u_l(\bvec{x}^*, j_1) = \Delta, u_l(\bvec{x}^*, j_2) = \Delta - c$, $u_f(\bvec{x}^*, j_1) = u_f(\bvec{x}^*, j_2) = \Delta$. Hence, the $\delta$-best response to $x^*$ is not unique, as $\BR_{\delta}(\bvec{x}^*) = \{j_1, j_2\}$. Moreover, the leader utility strictly improves by some constant gap $c$ if the follower does not follow the pessimistic tie-breaking rule, i.e., 
$$ \min_{j\in \BR_{\delta}(\bvec{x}^*) } u_{l}(\bvec{x}^* , j) < \max_{j\in \BR_{\delta}(\bvec{x}^*) } u_{l}(\bvec{x}^* , j) - c . $$
As we can set $c$ to be arbitrarily close to $\delta$ in the problem instance, we have
$$ \min_{j\in \BR_{\delta}(\bvec{x}^*) } u_{l}(\bvec{x}^* , j) < \max_{j\in \BR_{\delta}(\bvec{x}^*) } u_{l}(\bvec{x}^* , j) - \delta . $$

\begin{table}[h]
\begin{center}
\begin{tabular}{ |c|c|c|c| } 
 \hline
  $u_l,u_f$& $j_1$ & $j_2$ \\
 \hline
 $i_1$ &$0, \Delta$ & $0,0$ \\ 
 \hline
 $i_2$ & $\Delta, \Delta$ & $\Delta-c, \Delta$  \\
 \hline
 $i_3$ & $0, 0$ & $0, \Delta$  \\
 \hline
\end{tabular}
\end{center}
\caption{A class of game instances in which the $\delta$-RSE leader strategy does not have unique $\delta$-best response for any $0 < c < \delta < \Delta $.}
\label{table:nonunique-delta-br-set}
\end{table}
\qed

\subsection{Omitted Examples in the Proof of Theorem \ref{thm:delta-RSE-property}}
\label{sec:appendix-proof-section-property}

\subsubsection{Example for Property \ref{rse-property-1}}\label{sec:appendix-proof-section-property1}
We provide a Stackelberg game instance to show how the equality $u_{\textup{SSE}} = u_{\textup{RSE}}(0^+)$ fails to hold, if the non-degeneracy assumption $\Delta > 0$ condition is not satisfied. Consider the following game in Table \ref{table:1x2-noninducable}  where the leader only has a single action while the follower has two actions. It is easy to see that $\Delta = 0$ in this game. Clearly, the SSE leader utility would be $1$ which is induced by follower responding with action $j_2$. In contrast, we have $\BR_{0^+}(i_1) = \{j_1, j_2\}$ and by definition we have $u_{\textup{RSE}}(0^+) = 0$ which is induced by follower responding with action $j_1$.  
\begin{table*}[tbh]
\begin{center}
\begin{tabular}{ |c|c|c| } 
 \hline
  $u_l,u_f$& $j_1$ & $j_2$  \\
 \hline
 $i_1$ &$0, 1$ & $1,1$ \\
 \hline
\end{tabular}
\end{center}
\caption{A Stackelberg game instance whose SSE leader utility is different from $u_{\textup{RSE}}(0^+)$.}
\label{table:1x2-noninducable}
\end{table*}

However, note that $\Delta > 0$ is only a sufficient but not necessary condition in general game. More precisely, for the equality to be attainable, it is equivalent to whether the best response region of the SSE follower action has non-zero measure, which ensures the existence of strategy $\bvec{x}$ with $\BR_{0^+}(\bvec{x}) = \{j^*\}$ and so is the limit towards the SSE $(\bvec{x}^*, j^*)$.

\subsubsection{Examples for Property \ref{rse-property-3}}\label{sec:appendix-proof-section-property3}

\noindent\textbf{An example of continuous $u_{\textup{RSE}}(\delta)$. }\quad 
Table \ref{table:continuous-delta-utility} illustrates an instance whose inducibility gap $\Delta = 1$ and its $u_{\textup{RSE}}(\delta) = 
\begin{cases}
1 &\text{if $\delta \leq \epsilon$} \\ 
\frac{1-\delta}{1-\epsilon} &\text{$ \epsilon < \delta \leq 1$} \\
0 &\text{$ \delta > 1$}
\end{cases}$ is continuous for any $\delta > 0$.
\begin{table*}[tbh]
\begin{center}
        \begin{tabular}{|c|c|c|}
        \hline
        $u_{l}, u_{f}$ & $j_1$ & $j_2$ \\
        \hline
        $i_1$ & $1, \frac{1+\epsilon}{2}$ & $0, \frac{1-\epsilon}{2}$  \\ 
        \hline
        $i_2$ & $0, 0$ & $0, 1$ \\
        \hline
        $i_3$ & $0, 1$ & $0, 0$ \\
        \hline
        \end{tabular}
\end{center}
\caption{A Stackelberg game instance whose $u_{\textup{RSE}}(\delta)$ is continuous in $\delta$ ($\epsilon$ is any constant within $[0,1]$). }
\label{table:continuous-delta-utility}
\end{table*}

\begin{figure}[tbh]
    \centering
    \includegraphics[width= 0.8\textwidth]{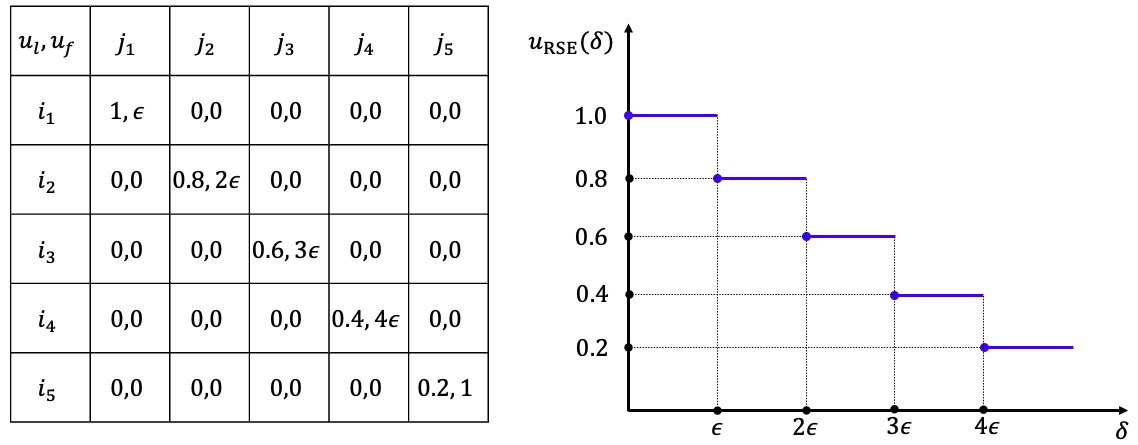}
    \caption{A Stackelberg game instance with $\Delta = \epsilon$, and the corresponding $u_{\textup{RSE}}(\delta)$ as a function of $\delta$.}%
    \label{fig:delta_utility}
\end{figure}

\noindent\textbf{An example of discontinuous for $u_{\textup{RSE}}(\delta)$ at $\delta \ge \Delta$.}\quad
Figure \ref{fig:delta_utility} illustrates an example  with discontinuous $u_{\textup{RSE}}(\delta)$. This directly showcases how $u_{\textup{RSE}}(\delta)$ \emph{may} be discontinuous at $\delta \geq \Delta$.

\subsection{Additional Discussions on Tie-breaking Rules and  Uniqueness of $\delta$-RSE} \label{sec:other-properties} 
It is well known that SSE is almost always unique, and has generically unique leader payoff regardless of the follower's tie-breaking rules~\citep{von2010leadership}. Hence, a natural question to ask is whether the $\delta$-RSE shares any of these properties. The answer to the first part of the question is clear: $\delta$-RSE is almost always unique in randomly generated game instances. This is because, as we fix the follower's pessimistic tie-breaking rule in Definition~\ref{def:delta-robust-SSE}, the conditions for optimization problem in Equation \eqref{eq:epsilon-robust-equilibrium} to admit multiple maximizers or minimizers have zero measure.  

For the second part of the question, it is obvious that in general different tie-breaking rules result in different optimal leader strategies, and thus different utilities,
$$ \max_{\bvec{x} \in \Delta^m}\max_{j\in \BR_{\delta}(\bvec{x}) } u_{l}(\bvec{x} , j) \neq \max_{\bvec{x} \in \Delta^m} \min_{j\in \BR_{\delta}(\bvec{x}) } u_{l}(\bvec{x} , j). $$ Instead, a more interesting  question is whether the $\delta$-RSE leader strategy $\bvec{x}^*$ has generically unique leader payoff if the follower could switch to different tie-breaking rules, i.e.,
$$ \max_{j\in \BR_{\delta}(\bvec{x}^*) } u_{l}(\bvec{x}^* , j) = \min_{j\in \BR_{\delta}(\bvec{x}^*) } u_{l}(\bvec{x}^* , j). $$
It is still obvious in the extreme cases where $\delta$ is larger than any utility difference between follower actions in the game, we have $\BR_{\delta}(\bvec{x}) = [m], \forall \bvec{x}$.
However, if we were to restrict to reasonably small $\delta$ such that $\delta < \Delta$, it becomes a more difficult question. Lemma \ref{lm:generic-payoff} shows that $\delta$-RSE do have generically unique leader payoff in this case when the number of leader actions $m=2$. However, this observation is no longer true in general games. 
In Proposition \ref{prop:tie-breaking}, we are able to construct a class of instances that the follower's response to the $\delta$-RSE leader strategy is not necessarily unique even in the case when $\delta$ is smaller than the inducibility gap $\Delta$. Hence, the follower's different tie-breaking rules could often result in radically different choices of the optimal leader strategy (with constant gaps among the corresponding leader payoffs). This is a shape contrast to the reckoning of SSE, where the follower's optimistic tie-breaking is almost without loss of generality, since the leader (in generic games) could improve his payoff by changing his strategy slightly so that the desired reply is unique in the equilibrium.  
As a result, we cannot apply the standard approaches used for SSE for the computation of $\delta$-RSE.  These differences are indeed rooted in the design of the two solution concepts: SSE serves as a standard solution concept (in generic games) where the follower's tie-breaking rule should not matter, whereas it is natural for $\delta$-RSE, a robust solution concept, to focus on the worst case scenario and assume pessimistic tie-breaking rule.

\begin{lemma}\label{lm:generic-payoff}
In any game with $m=2$, $\Delta > 0, \delta \in (0, \Delta)$, at least one of its $\delta$-RSE leader strategy $\bvec{x}^*$ satisfies
$$ \max_{j\in \BR_{\delta}(\bvec{x}^*) } u_{l}(\bvec{x}^* , j) = \min_{j\in \BR_{\delta}(\bvec{x}^*) } u_{l}(\bvec{x}^* , j). $$
\end{lemma}

\noindent\textit{Proof.}\quad
Pick one of $\delta$-RSE leader strategy $\bvec{x}^*$ in the game. The lemma clearly holds, when the $\delta$-RSE leader strategy have $|\BR_{\delta}(\bvec{x}^*)|=1$. A key observation of this proof is the characterizations in Lemma \ref{lm:br-characterization} such that it suffices to show that if the unique $\delta$-RSE leader strategy $\bvec{x}^*$ satisfies $\BR_{\delta}(\bvec{x}^*) = \{j,j'\}$ for some $j,j' \in [n]$, then $u_l(\bvec{x}^*, j) =  u_l(\bvec{x}^*, j')$. We prove this statement by contradiction:

Suppose there is a unique $\delta$-RSE leader strategy $\bvec{x}^*$ where $\BR_{\delta}(\bvec{x}^*) = \{j,j'\}$, $u_l(\bvec{x}^*, j) < u_l(\bvec{x}^*, j')$. Under the pessimistic tie-breaking, the follower will respond with $j$. We first rule out the case that $u_l(\cdot, j) $ is a constant function, because we could find another leader strategy $\bvec{x}$ such that $u_l(\bvec{x}^*, j) = u_l(\bvec{x}, j) $ and $\BR_{\delta}(\bvec{x}) = \{j\}$, due to Property \ref{claim:m=1} in Lemma~\ref{lm:br-characterization}. 

Let $\bvec{x}^L, \bvec{x}^R$ be the two strategy on the boundary $\cX(\delta; \{j,j'\})$. According to Property \ref{claim:m=2} in Lemma \ref{lm:br-characterization}, we can without loss of generality assume $\BR_{\delta}(\bvec{x}^L) = \{j\},  \BR_{\delta}(\bvec{x}^R) = \{j'\}$. We argue that $\bvec{x}^*$ is not an RSE leader strategy, since $ \max \left\{ u_l(\bvec{x}^L, j), u_l(\bvec{x}^R, j') \right\}  > u_l(\bvec{x}^*, j) $. Notice that, if we move $\bvec{x}^*$ towards one of the two open boundary points of $\cX(\delta; \{j,j'\})$ where $u_l(\cdot, j)$ is increasing, two possible cases can happen: 1) It reaches the point where $u_l(\bvec{x}^*, j) = u_l(\bvec{x}^*, j')$, and it violates the assumption. 2) We can set $\bvec{x}^*$ to be arbitrarily close to the right of $\bvec{x}^L$ or to the left of $\bvec{x}^R$. If it is to the right of $\bvec{x}^L$, then clearly $u_l(\bvec{x}^L, j) > u_l(\bvec{x}^*, j)$. 
If it is to the left of $\bvec{x}^R$, then we have $u_l(\bvec{x}^R, j') = \lim_{\bvec{x}^* \to \bvec{x}^R} u_l(\bvec{x}^*, j') > \lim_{\bvec{x}^* \to \bvec{x}^R} u_l(\bvec{x}^*, j)$. Hence, $\bvec{x}^*$ is not an RSE leader strategy. 

\qed

\begin{lemma}[A Characterization of $\delta$-Best Response Regions]\label{lm:br-characterization}
Denote by $\cX(\delta;S) := \{\bvec{x} \in \Delta^m | \BR_{\delta}(x) = S  \}$, the $\delta$-best response region for a subset of follower actions $S \subseteq [n]$. $\cX(\delta;S) $ is a (possibly empty) convex set and satisfies the following properties when $\Delta > 0, \delta \in (0, \Delta)$:
\begin{enumerate}
    \item\label{claim:m=1} If $|S| = 1$, then $\cX(\delta;S) \neq \varnothing$ and $\cX(\delta;S) \supset \cX(\delta';S)$,  $\forall \delta' \in (\delta, \Delta)$.
    \item\label{claim:m>2} If $m=2$ and $|S| \geq 3$, then $\cX(\delta;S) = \varnothing$. 
    \item\label{claim:m=2} If  $m=2$ and $|S| = 2$, then  $\cX(\delta;S) $ is an nonempty open set for any $S=\{j,j'\}$ that $\exists \bvec{x}\in \Delta^m$ with $u_f(\bvec{x}, j)=u_f(\bvec{x}, j')$, the boundary of $\cX(\delta;S) $ contains exactly two points $\bvec{x}^L, \bvec{x}^R$ with $\BR_{\delta}(\bvec{x}^L) = \{ j\},\BR_{\delta}(\bvec{x}^R) =  \{ j'\}$, and moreover, $ \cX(\delta;S) \subset \cX(\delta';S)$, $\forall \delta' \in (\delta, \Delta)$.
\end{enumerate}
\end{lemma}
\noindent\textit{Proof.}\quad
Since $\BR_{\delta}(\bvec{x})$ definition in Equation \eqref{def:delta-robust-set} specifies a set of convex constraints,  $\cX(\delta;S) $ forms a convex set as a result of finite intersection of convex sets. Below, we prove each one of the properties:

\paragraph{\bf Property \ref{claim:m=1}}
It holds universally for any $m$. The first part can be directly verified using the definition of $\Delta$. That is, $\forall j\in [n]$, there exists $\bvec{x}\in \Delta^m $ such that $u_f(\bvec{x}, j) - u_f(\bvec{x}, j') \geq \Delta > \delta, \forall j\neq j'$, which means that $\cX(\delta; \{j\}) \neq \varnothing$. For the second part, observe that the best response region of any action $j\in [n]$ can be written as $\cX(\delta;{j}) = \{\bvec{x} \in \Delta^m | u_f(\bvec{x}, j)  \geq  \max_{j'\neq j}u_f(\bvec{x}, j') + \delta  \}$. It is thus clear that $\cX(\delta;S) \supseteq \cX(\delta';S) $, $\forall \delta' \in (\delta, \Delta)$.  To go from $\supseteq $ to $\supset$, it suffices to show that  $\cX(\delta;S) \setminus \cX(\delta';S) $ is not empty for any $\delta'-\delta > 0$. That is, for any $\delta \in (0, \Delta)$, there exists $\bvec{x}^\delta_j \in \Delta^m$ such that $u_f(\bvec{x}^\delta_j, j)  =  \max_{j'\neq j}u_f(\bvec{x}^\delta_j, j') + \delta$. This is because we have $\bvec{x}_1, \bvec{x}_2 \in \Delta^m$ such that  $u_f(\bvec{x}_1, j)  \geq  \max_{j'\neq j}u_f(\bvec{x}_1, j') + \Delta$ and $u_f(\bvec{x}_2, j)  \geq  \max_{j'\neq j}u_f(\bvec{x}_2, j') $ as $\Delta > 0$. So $\bvec{x}^\delta_j \in \Delta^m$ as a convex combination of $\bvec{x}_1, \bvec{x}_2$. Hence, $\cX(\delta;S) \supset \cX(\delta';S)$,  $\forall \delta' \in (\delta, \Delta)$.

\paragraph{\bf Property \ref{claim:m>2}}
We prove this by contradiction. Suppose there is such $\bvec{x} \in \cX(\delta; S)$ with $|S| > 2$.  Pick three follower actions, say $j_1,j_2,j_3$ with the highest follower utility at $\bvec{x}$, and we can assume without loss of generality that,
$$ u_f(\bvec{x}, j_1) \geq u_f(\bvec{x}, j_2 ) \geq u_f(\bvec{x}, j_3 ) > u_f(\bvec{x}, j_1) - \delta. $$ 
We first argue that the above inequalities must be strict. That is, the following cases cannot hold:
\begin{enumerate}
    \item\label{case-1} $u_f(\bvec{x}, j_1) = u_f(\bvec{x}, j_2 ) = u_f(\bvec{x}, j_3 ) > u_f(\bvec{x}, j_1)-\delta.$ 
    \item\label{case-2} $u_f(\bvec{x}, j_1) > u_f(\bvec{x}, j_2 ) = u_f(\bvec{x}, j_3 ) > u_f(\bvec{x}, j_1)-\delta.$
    \item\label{case-3} $u_f(\bvec{x}, j_1) = u_f(\bvec{x}, j_2 ) > u_f(\bvec{x}, j_3 ) > u_f(\bvec{x}, j_1) - \delta. $ 
\end{enumerate}

For the analysis  below, we consider the gradient functions on the follower's utility over action $j_1, j_2, j_3$, as $f_1 = \frac{d [ u_f( (x, 1-x) , j_1)] }{d {x}}, f_2 = \frac{d [ u_f( (x, 1-x) , j_2)] }{d {x}}, f_3 = \frac{d [ u_f( (x, 1-x) , j_3)] }{d {x}} $. Since the utility functions are linear, we can treat $f_1, f_2, f_3$ as constants. Moreover, $f_1 \neq f_2 \neq f_3$, or it would indicate constant utility gap between two actions, contradicting with $\Delta > 0$. By symmetry, assume without loss of generality $f_1 - f_2 < 0$.

Case \ref{case-1}: Since $f_1 \neq f_2 \neq f_3$, there exists a strict ordering among them. Pick any ordering, say $f_1 < f_2 < f_3$. Then $u_f(\bvec{x}', j_2 ) < u_f(\bvec{x}', j_1 )$, for any $\bvec{x}'$ on the left of $\bvec{x}$, and $u_f(\bvec{x}', j_2 ) < u_f(\bvec{x}', j_3 )$, for any $\bvec{x}'$ on the right of $\bvec{x}$, which contradict with $\Delta > 0$.

Case \ref{case-2}: Note that $f_1 - f_2$ (or $f_1-f_3$) captures the change of utility difference between $j_1$ and $j_2$ (or $j_3$).  
Since $f_1 \neq f_2 \neq f_3$, it is only possible that $(f_1 - f_2)(f_1 - f_3) > 0$ or $(f_1 - f_2)(f_1 - f_3) < 0$. If $(f_1 - f_2)(f_1 - f_3) < 0$, the maximum utility difference between $j_1$ and $j_2, j_3$ is no more than $\delta$, which contradict with $\Delta > \delta$. If $(f_1 - f_2)(f_1 - f_3) > 0$ and $f_1 - f_2 < 0$ by assumption, we  have $f_1 -f_3 <0$.  Pick any ordering between $f_2, f_3$, say $f_2 < f_3$. Then, $j_2$ can never be the follower's best response: for any leader strategy on the left of $\bvec{x}$, the follower's utility of $j_2, j_3$ must be worse than that of $j_1$; for any leader strategy on the right of $\bvec{x}$, the follower's utility of $j_2$ must be worse than that of $j_3$. This again contradicts $\Delta > 0$.

Case \ref{case-3}: 
Since $f_1 - f_2 < 0$, it is only possible that $f_1-f_3 < 0$ (or $f_1-f_3 > 0$ and  $f_2-f_3 > 0$). %
In this case, the follower's utility difference between $j_3$ and $j_1$ (or $j_2$) will always be less $\delta$, and thus, $\cX(\delta;\{j_1\})$ (or $\cX(\delta;\{j_2\})$ ) will be empty, contradicting Property \ref{claim:m=1}.

It now remains to show that even when the inequalities are not strict, the case $u_f(\bvec{x}, j_1) > u_f(\bvec{x}, j_2 ) > u_f(\bvec{x}, j_3 ) > u_f(\bvec{x}, j_1) - \delta $ is also impossible. Again since $f_1 - f_2 < 0$, it is only possible that $f_1-f_3 < 0$, (or $f_1-f_3 < 0$ and  $f_2-f_3 > 0$): 
\begin{itemize}
    \item If $f_1-f_3 > 0$ and $f_2-f_3 > 0$, then there exists a leader strategy $\bvec{x}'$ on the left of $\bvec{x}$ such that 
    $$  u_f(\bvec{x}', j_1) \geq u_f(\bvec{x}', j_2 ) = u_f(\bvec{x}', j_3 ) >  u_f(\bvec{x}', j_1) - \delta, $$
    which reduces to Case \ref{case-1} or \ref{case-2} above and reach the contradiction. 
    \item If $f_1-f_3 < 0$, then there exists a leader strategy $\bvec{x}'$ on the right of $\bvec{x}$ such that 
    $$u_f(\bvec{x}, j_1) = u_f(\bvec{x}', j_2 ) \geq u_f(\bvec{x}', j_3 ) > u_f(\bvec{x}', j_1) - \delta,$$
     which reduces to Case \ref{case-1} or \ref{case-3} above and reach the contradiction. 
\end{itemize}

\paragraph{\bf Property \ref{claim:m=2}}
The first part of the property hinges on the observation that, when $m=2$, for any $j,j'\in [n]$, if there exists $\bvec{x}^{j,j'} \in \Delta^m$ such that $\BR(\bvec{x}^{j,j'}) = \{j,j'\}$, then $\bvec{x}^{j,j'} \in \cX(\delta; \{j,j'\})$ for any $\delta \in (0, \Delta)$. This is because $\bvec{x}^{j,j'} \not\in \cX(\delta; S), \forall |S| = 1$ by definition, or $|S| > 2$ by Property \ref{claim:m>2}. Given this observation, we prove by contradiction that $\cX(\delta; \{j,j'\})$ must only contain points from the interior of $\cX(0;{j}), \cX(0;{j'})$, except for $\bvec{x}^{j,j'}$. Suppose there exists another strategy $\bvec{x} \in \cX(\delta; \{j,j'\})$ such that $\BR(\bvec{x}) = \{ j'' \}$ and $j'' \neq j, j'$. Assume WLOG a left to right ordering of $\cX(0;{j}), \cX(0;\{j'\}), \cX(0;\{j''\})$, i.e., the best response region of $j, j', j''$, respectively, in the 1-dimensional simplex. By convexity, this implies that $\cX(\delta; \{j,j'\}) \supseteq \cX(0; \{j'\} \supseteq \cX(\delta; \{j'\})$. Since $\cX(\delta; \{j,j'\})\cap \cX(\delta; \{j'\})=\varnothing $, we get $ \cX(\delta; \{j'\})=\varnothing$, contradicting Property~\ref{claim:m=1}. %
Using the same argument, $\cX(\delta; \{j,j'\})$ cannot contain points from the boundary of $\cX(0;{j}), \cX(0;{j'})$, except for $\bvec{x}^{j,j'}$.
In addition, the two boundary point $\bvec{x}^L, \bvec{x}^R$ of $\cX(\delta; \{j,j'\})$  must satisfy $u_f(\bvec{x}^L, j )  - u_f(\bvec{x}^L, j' ) = \delta $ and $u_f(\bvec{x}^R, j ) - u_f(\bvec{x}^R, j' ) = -\delta $ and thus $\bvec{x}^L, \bvec{x}^R \not\in \cX(\delta; \{j,j'\})$, $\cX(\delta; \{j,j'\})$ is an open set. Suppose if $u_f(\bvec{x}^L, j )  - u_f(\bvec{x}^L, j' ) = \delta' $. Clearly, if $\delta' > \delta$, the $\bvec{x}^L$ must not be on the boundary of $\cX(\delta; \{j,j'\})$ by definition. The case where $\delta' < \delta$ is also impossible, because this means $\bvec{x}^L$ is on the left boundary of $\bvec{x}$, or moving $\bvec{x}^L$ leftward increases the utility gap $\delta'$. Similar argument holds for $u_f(\bvec{x}^R, j ) - u_f(\bvec{x}^R, j' ) = -\delta $. 

The proof for the second part of the property now becomes straightforward. Consider $\delta' = \delta + \epsilon$ for any $\epsilon > 0, \delta' < \Delta$. We can verify by definition that $\cX(\delta; \{j,j'\}) \subseteq \cX(\delta'; \{j,j'\})$. That is, for any $\bvec{x} \in \cX(\delta; \{j,j'\})$, $ |u_f(\bvec{x}, j') - u_f(\bvec{x}, j)| < \delta < \delta - \epsilon$. 
Moreover, observe that for the two points on the open boundary of $\cX(\delta; \{j,j'\})$, we have $\bvec{x}^R, \bvec{x}^L \in \cX(\delta+\epsilon; \{j,j'\})$. Therefore, $\cX(\delta; \{j,j'\}) \subset \cX(\delta'; \{j,j'\})$.
\qed

\subsection{Proof of Proposition \ref{prop:rse-convexity}} \label{appendix:rse-convex}
\noindent\textit{Proof.}\quad
For the case when $m=2$. Let $ U(\delta;S) := \max_{\bvec{x} \in \cX(\delta;S) } \min_{j\in S} u_l(\bvec{x}, j)$.  Then, $u_{\textup{RSE}}(\delta) = \max_{S\subseteq [n]} U(\delta;S) = \max_{S\subseteq [n], |S|\leq 2} U(\delta;S) $, by Lemma \ref{lm:br-characterization}. 
We first prove a claim that if $U(\delta; S)$ is linearly for all $|S|=1$, then $u_{\textup{RSE}}(\delta)$ is convex at $\delta \in (0, \Delta)$. 

From Lemma \ref{lm:br-characterization}, we know $U(\delta;S)$ is non-increasing in $\delta$ when $|S|=1$, and is non-decreasing in $\delta$ when $|S|=2$. 
This, along with Lemma \ref{lm:generic-payoff}, implies that if for some $\delta_0 \in (0, \Delta)$, $\delta_0$-RSE leader strategy $\bvec{x}^* \in  \cX(\delta;S)$ with $|S|=2$, then $\bvec{x}^*$ is the $\delta$-RSE for all $\delta \in [\delta_0, \Delta)$ --- otherwise, it breaks the continuity of $u_{\textup{RSE}}(\delta)$ when $\delta \in (0, \Delta)$.
Hence, if such $\delta_0$ exists, then we have
$$
u_{\textup{RSE}}(\delta) = 
\begin{cases}
\max_{S\subseteq [n], |S| = 1} U(\delta;S) & \delta \in (0, \delta_0) \\
u_{\textup{RSE}}(\delta_0) & \delta \in [\delta_0, \Delta)
\end{cases},
$$
where $u_{\textup{RSE}}(\delta)$ is convex non-increasing when $\delta \in (0, \delta_0)$ and is constant when $\delta \in [\delta_0, \Delta)$ --- its gradient is non-decreasing and thus $u_{\textup{RSE}}(\delta)$ is convex.  Meanwhile, if such $\delta_0$ does not exist, $u_{\textup{RSE}}(\delta) = \max_{S\subseteq [n], |S| = 1} U(\delta;S)$ is clearly convex. 

It only remains to show that $U(\delta; S)$ is linearly for all $|S|=1$. To see this, we can explicitly derive its gradient. Let $S = \{j_1\}$ and the leader utility gradient $g_1 = \frac{u_l((x,1-x), j_1)}{d x}$, follower utility gradient $f_i = \frac{u_f((x,1-x), j_i)}{d x}, \forall i\in [n]$. We assume WLOG $g_1 < 0$; otherwise, if $g_1 = 0$, $U(\delta; S)$ is a constant function and thus linear.
The Lagrangian of $U(\delta; \{j_1\})$  is $L(\delta, x) = g_1 x - \sum_{i=2}^{n}  \lambda_i \left[ (f_1 - f_i)x - \delta \right]  $, so its gradient function can be simplified using the envelope theorem as,
$$ 
\frac{\partial U(\delta; \{j_1\})}{\partial \delta} = \frac{\partial L(\delta, x) }{\partial \delta} = \sum_{i=2}^{n} \lambda_i.
$$
So it only remains to determine $\lambda_i$.  First, if no constraint is tight, then $\sum_{i=2}^{n} \lambda_i = 0$ and $U(\delta; \{j_1\})$ is linear. Second, observe that, since the optimal solution $x$ lies in a 1-dimensional space, there can be at most one constraint that is tight for any $\delta \in (0, \Delta)$, i.e., $ u_f(\bvec{x}, \{j_1\}) - u_f(\bvec{x}, \{j_i\}) = \delta $ for some $j_i \neq j_1$. In this case, the gradient is constant $\frac{\partial U(\delta; \{j_1\})}{\partial \delta} = \lambda_i = \frac{g_1}{f_i - f_1}$ and $U(\delta; \{j_1\})$ is linear.

For the case when $m>2$, we prove by construction. Consider a game with $m=3, n=2$ and utility matrices specified in Table~\ref{table:nonConvex} where the parameter $\Delta \in (0, \frac{1}{2})$ and $c \in (0, 1)$. 

\begin{table}[h]
\begin{center}
\begin{tabular}{ |c|c|c|c| } 
 \hline
  $u_l,u_f$& $j_1$ & $j_2$ \\
 \hline
 $i_1$ &$0, 1$ & $c, \Delta$ \\ 
 \hline
 $i_2$ & $\sfrac{1}{2}, 0$ & $c, \Delta$  \\
 \hline
 $i_3$ & $1, \sfrac{1}{2}$ & $c, \Delta$  \\
 \hline
\end{tabular}
\end{center}
\caption{An instance where $u_{\textup{RSE}}(\delta)$ is neither convex nor concave for $\Delta \in (0, \frac{1}{2}), c \in (0, 1)$.}
\label{table:nonConvex}
\end{table}

\begin{figure}
    \centering
    \includegraphics[width=0.24\textwidth]{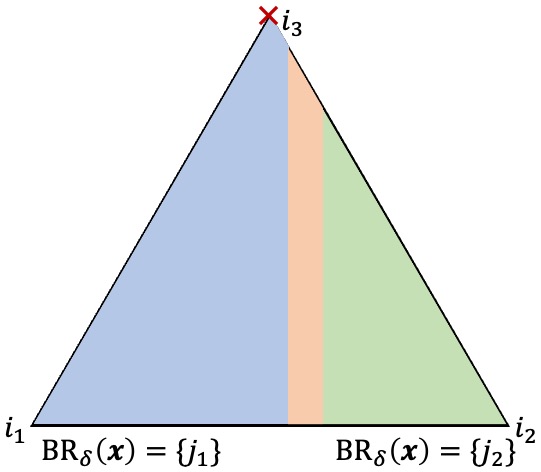}
    \includegraphics[width=0.24\textwidth]{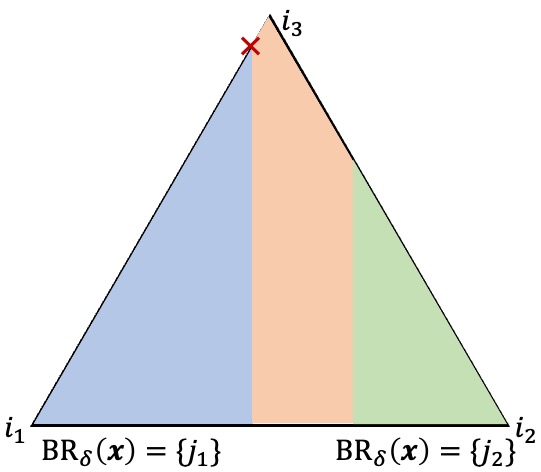}
    \includegraphics[width=0.24\textwidth]{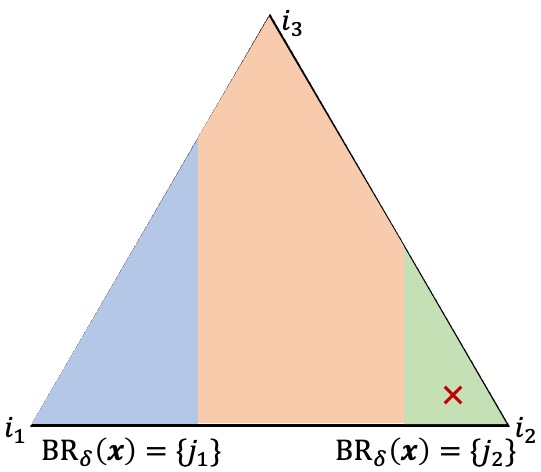}
    \includegraphics[width=0.24\textwidth]{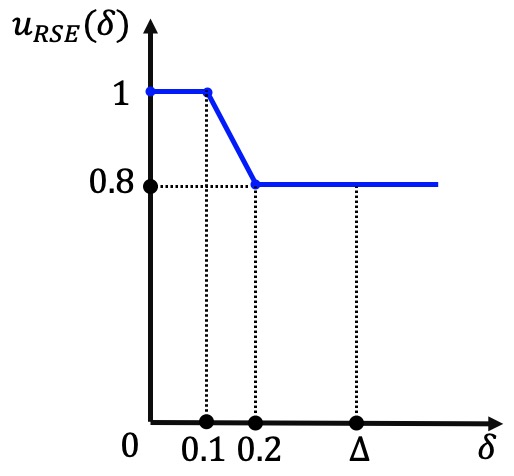}
    \caption{An illustration of $\delta$-RSE strategy $\bvec{x}^*$ (highlighted with red marks) in the leader's strategy simplex and the plot of the function $u_{\textup{RSE}}(\delta)$, when $\Delta=0.4, c=0.8$. Each one of the first plots represents a linear segment of $u_{\textup{RSE}}(\delta)$.   }
    \label{fig:u-rse-shape}
\end{figure}

We start with some observations on this game: 1) follower always have utility $\Delta$ for any leader strategy; 2) Since $c < 1$, the leader utility is maximized at the action profile $(i_3, j_1)$. We now determine the $\delta$-RSE at different interval of $\delta$, see Figure \ref{fig:u-rse-shape} for an illustration:
\begin{itemize}
    \item For $\delta \in (0, \frac12-\Delta]$, the $\delta$-RSE leader strategy $\bvec{x}^* = (0,0,1)$, as $\BR_{\delta}(\bvec{x}^*) = \{j_1\}$ and the leader receives utility $1$ that are strictly better than any other possible leader/follower action profile. 
    \item For $\delta \in (\frac12-\Delta, 1-\frac{c}{2} - \Delta]$, the $\delta$-RSE leader strategy $\bvec{x}^* = (\epsilon,0,1-\epsilon)$, where $\epsilon = 2\Delta + 2\delta - 1$. We can verify that $\BR_{\delta}(\bvec{x}^*) = \{j_1\}$, since $\frac{1}{2}(1-\epsilon) + \epsilon \geq \Delta + \delta$ and the leader receives utility $1-\epsilon=2 - 2\Delta - 2\delta$ that are no worse than $c$, which is the leader utility if the follower ever responds with $j_2$.
    \item For $\delta \in (1-\frac{c}{2} - \Delta, \infty)$, the $\delta$-RSE leader strategy $\bvec{x}^*=(0,1,0)$. This is because leader cannot achieve a utility more than $c$, and thus the $\delta$-RSE leader strategy $\bvec{x}^* $ only needs to satisfy that $\BR(\bvec{x}^*) = \{j_2\}$.
\end{itemize}
Therefore, it is clear that $u_{\textup{RSE}}(\delta)$ in this instance is neither convex nor concave.

\section{Omitted Proofs in Section \ref{sec:computational}}\label{appendix-computational-section}

\subsection{Proof of Corollary \ref{obv:efficent-rse}}\label{appendix-computational-obv}
\textit{Proof.}\quad
According to the proof of Property (2) in Theorem \ref{thm:delta-RSE-property}, we can construct a specific leader strategy $\hat{\bvec{x}} = (1 - \frac{\delta}{\Delta}) \bvec{x}^* + \frac{\delta}{\Delta} \bvec{x}^{j^{*}}$ whose leader utility against a $\delta$-rational follower is at least $(1-\frac{\delta}{\Delta})u_{\textup{SSE}}$.  
Recall that $\langle \bvec{x}^*, j^*\rangle$ is the SSE of the game, and $\bvec{x}^{j^{*}}$ is the strategy such that $u_f(\bvec{x}^{j^{*}}, j^*) \geq u_f(\bvec{x}^{j^{*}}, j') + \Delta$ for all $j' \neq j^*$. 
It is easy to see that its utility $(1-\frac{\delta}{\Delta}) u_{\textup{SSE}} \geq (1-\frac{\delta}{\Delta}) u_{\textup{RSE}}(\delta)  \geq u_{\textup{RSE}}(\delta) - \frac{\delta}{\Delta}$, which is $\frac{\delta}{\Delta}$-optimal for the $\delta$-RSE leader utility. 

It remains to show this construction is computationally efficient: it is well-known that SSE $( \bvec{x}^*, j^*)$ can be computed by solving $n$ linear programs \cite{conitzer2006computing}. What's more, finding $\bvec{x}^{j^{*}}$ requires solving the following linear program:
\begin{equation}\label{eq:find_strict_strategy}
    \begin{split}
        \max_{\bvec{x}} \quad &u_l(\bvec{x}, j^*) \\
        \text{subject to} \quad &u_f(\bvec{x}, j^*) \geq u_f(\bvec{x}, j') + \Delta, \forall j' \neq j^*
    \end{split}
\end{equation}
Thus, constructing $\hat{\bvec{x}}$ requires solving $n+1$ linear programs, and the total time complexity is $O(\poly(m,n))$.
\qed

\section{Omitted Proofs in Section \ref{sec:learnability}}\label{sec:appendix-learning-section}

\subsection{Proof of Lemma \ref{lm:delta-eps-trade-off}}
\label{sec:proof-delta-eps-trade-off}

Let $\BR_{\delta}(\bvec{x}, u_f) $ denote the set of $\delta$-optimal response(s) of $\bvec{x}$ for follower with utility function $u_f$.
We start by showing that $\forall \bvec{x}\in \Delta^m, \BR_{\delta+2\epsilon}(\bvec{x}, \tilde{u}_f) \supseteq \BR_{\delta}(\bvec{x}, u_f)$, and it suffices to argue that for any $j \in [n]$, if  $ j \not\in \BR_{\delta+2\epsilon}(\bvec{x}, \tilde{u}_f)$, then  $ j \not\in \BR_{\delta}(\bvec{x}, u_f)$. That is, given $j \not\in \BR_{\delta+2\epsilon}(x, \tilde{u}_f)$, we know there exists $j'$ such that $\tilde{u}_f(\bvec{x}, j) - \tilde{u}_f(\bvec{x}, j') \leq - \delta - 2\epsilon $. Reusing the fact that for any $\norm{\tilde{u} - u}_{\infty} \leq \epsilon$, $\bvec{x} \in \Delta^m$, $|\tilde{u}(\bvec{x},j) - u(\bvec{x},j)| \leq \epsilon $, we have ${u}_f(\bvec{x}, j) - {u}_f( \bvec{x}, j') \leq \tilde{u}_f(\bvec{x}, j) - \tilde{u}_f(\bvec{x}, j') + 2\epsilon \leq -\delta $ and thus $j \not\in \BR_{\delta}(\bvec{x}, u_f)$.

Since $\BR_{\delta+2\epsilon}(\bvec{x}, \tilde{u}_f) \supseteq \BR_{\delta}(\bvec{x}, u_f)$, we have by definition,
$$ V(\bvec{x} ; u_l, u_f, \delta) 
= \min_{j\in \BR_{\delta}(\bvec{x}, {u}_f)} u_l(\bvec{x}, j) 
\geq \min_{j\in \BR_{\delta+2\epsilon}(\bvec{x}, \tilde{u}_f)} u_l(\bvec{x}, j) 
 = V(\bvec{x} ; u_l, \tilde{u}_f, \delta + 2\epsilon). $$
 
This leads to the following inequalities,
$$ V^*(u_l, \tilde{u}_f, \delta + 2\epsilon) = V(\bvec{x}_1; u_l, \tilde{u}_f, \delta + 2\epsilon)
\leq V(\bvec{x}_1; u_l, u_f, \delta ) \leq V(\bvec{x}_2; u_l, u_f, \delta ) = V^*( u_l, u_f, \delta),   $$
where $\bvec{x}_1$ is $(\delta+ 2\epsilon)$-RSE strategy of Stackelberg game $u_l, \tilde{u}_f$, and $\bvec{x}_2$ is $\delta$-RSE strategy of Stackelberg game $u_l, u_f$.

\subsection{Proof of Proposition \ref{prop:learningwithgap}}\label{appendix-learning-prop}
\textit{Proof.}\quad
We divide the proof into two parts: $\delta < \Delta$ or $\delta \geq \Delta$. The proofs for two parts are similar but require different instance construction. 

\smallskip
\noindent\textbf{$\Omega( 1/\sqrt{T} )$ lower bound instance when $\delta < \Delta$:} 

Table \ref{table:hardness_learning_new} illustrates a game in which the inducibility gap is $\Delta$. Suppose $\Delta > \delta$. Consider the following two Stackelberg games $G_1$, $G_2$ where $r_{l}(i,j) \sim \text{Bern}(u_{l}(i,j))$ and $r_{f}(i,j) \sim \text{Bern}(u_{f}(i,j))$, with mean values shown in Table \ref{table:hardness_learning_new} and $\epsilon \in [0,1]$ is a parameter to be determined.

\begin{table*}
\begin{center}
        \begin{tabular}{|c|c|c|}
        \hline
        $u_{l}, u_{f}$ & $j_1$ & $j_2$ \\
        \hline
        $i_1$ & $1, \frac{1+\epsilon}{2} + \delta$ & $0, \frac{1-\epsilon}{2}$  \\ 
        \hline
        $i_2$ & $0, 0$ & $0, \Delta$ \\
        \hline
        $i_3$ & $0, \Delta$ & $0, 0$ \\
        \hline
        \end{tabular}
        \hspace{15mm}
        \begin{tabular}{|c|c|c|}
        \hline
        $u_{l}, u_{f}$ & $j_1$ & $j_2$\\
        \hline
        $i_1$ & $1, \frac{1-\epsilon}{2} + \delta$ & $0, \frac{1+\epsilon}{2}$\\
        \hline
        $i_2$ & $0, 0$ & $0, \Delta$\\
        \hline
        $i_3$ & $0, \Delta$ & $0, 0$\\
        \hline
        \end{tabular}
\end{center}
\vspace{2mm}
\caption{Example instances $G_1=\{u_l, u_f\}$ (left) and $G_2=\{u_l, u_f\}$ (right). Each table represents a Stackelberg game $u_l, u_f \in \mathbb{R}^{3\times 2}$ with inducibility gap $\Delta$.}
\label{table:hardness_learning_new}
\end{table*}

In Stackelberg game $G_1$, the leader strategy in $\delta$-RSE can be computed as 
\[\bvec{x}^*_1 = \argmax_{\bvec{x} \in \Delta^{m}} \min_{j \in \BR_{\delta}(\bvec{x})}  u_l(\bvec{x}, j) = (1, 0, 0).\]
That is, in the $\delta$-RSE of Stackelberg game $G_1$, the leader plays pure strategy $i_1$ and $\BR_\delta(i_1) = \{j_1\}$. As a result, we have $u_{l}^*(\delta) = 1$ for $G_1$. 

On the other hand, for Stackelberg game $G_2$, note that $u_{f}(i_1, j_1) - u_{f}(i_1, j_2) = \delta - \epsilon < \delta$. Thus, $j_2$ is included in the follower's $\delta$-optimal  response set if the leader plays pure strategy $i_1$, i.e. the $\bvec{x}^*_1$ for the other game $G_1$. Consequently, the leader receives $0$ utility if the leader plays pure strategy $i_1$ in the Stackelberg game $G_2$. However, the leader can play a mixed strategy $\bvec{x}^*_2$ that includes both pure actions $i_1$ and $i_3$ to increase the follower's utility difference between responding with $j_1$ and $j_2$. When the probability of playing $i_3$ is high enough, $\bvec{x}^*_2$ can make $u_{f}(\bvec{x}^*_2, j_1)- u_{f}(\bvec{x}^*_2, j_2) = \delta$. Specifically, let $\bvec{x}^*_2 = (p, 0, 1-p)$, then we have the following constraint on $p$ in order to exclude $j_2$ from the $\delta$-optimal response set:
\begin{equation}\label{eq:mixing_prob}
    p\cdot\left(\frac{1-\epsilon}{2} + \delta\right) + (1-p)\cdot \Delta - p \cdot \frac{1+\epsilon}{2} = \delta
\end{equation}
which makes $p = \frac{\Delta - \delta}{\Delta - \delta - \epsilon}$. As a result, for the Stackelberg game $G_2$, we have $\BR_{\delta}(\bvec{x}^*_2) = \{j_1\}$ and $u_{\textup{RSE}}(\delta) = u_{l}(\bvec{x}^*_2, j_1) = \frac{\Delta - \delta}{\Delta - \delta - \epsilon}$.

Therefore, if the learning algorithm cannot distinguish the above two Stackelberg games with finite samples and makes a wrong estimation, the outputted leader strategy will suffer an error of at least $\frac{\epsilon}{\Delta - \delta + \epsilon}$. Specifically, the following incorrect estimations can happen:
\begin{enumerate}
    \item If the true game is $G_1$, but the learning algorithm estimated the game as $G_2$ and outputs leader strategy as $\hat{\bvec{x}} = \bvec{x}^*_2$. Then the leader utility of playing $\bvec{x}^*_2$ will be $\frac{\Delta - \delta}{\Delta - \delta + \epsilon}$ while $u_{\textup{RSE}}(\delta) = 1$, this has the learning error of $\frac{\epsilon}{\Delta - \delta + \epsilon}$.
    \item If the true game is $G_2$, but the learning algorithm estimated the game as $G_1$ and outputs leader strategy as $\hat{\bvec{x}} = \bvec{x}^*_1$. Then the leader utility of playing $\bvec{x}^*_1$ will be $0$ while $u_{\textup{RSE}}(\delta) = \frac{\Delta - \delta}{\Delta - \delta + \epsilon}$, and this has the learning error of $\frac{\Delta - \delta}{\Delta - \delta + \epsilon}$ which is even larger than $\frac{\epsilon}{\Delta - \delta + \epsilon}$.
\end{enumerate}
Consequently, if the learning algorithm can output a leader strategy $\hat{\bvec{x}}$  that violates equation \eqref{eq:observation2}, i.e., 
$$\min_{j \in \BR_{\delta}(\hat{\bvec{x}})} u_l(\hat{\bvec{x}}, j) > u_{\textup{RSE}}(\delta) - \frac{\epsilon}{\Delta - \delta + \epsilon},$$ 
then for the game $G_1$ we have $u_{l}(\hat{\bvec{x}}, j) > \frac{\Delta - \delta}{\Delta - \delta + \epsilon}$,  while for the game $G_2$ we have $u_{\textup{RSE}}(\delta) = \frac{\Delta - \delta}{\Delta - \delta + \epsilon}$. In other words, the learning algorithm can distinguish  $G_1$ from  $G_2$ with $T$ samples.

Therefore, consider the learning algorithm that samples $r_f(i, j)$ for $T$ times, and the goal is to identify if $u_f \in G_1$ or $u_f \in G_2$. We prove the proposition by contradiction. Suppose Proposition \ref{prop:learningwithgap} is not correct, then with $T$ samples, we can  identify if $u_f \in G_1$ or $u_f \in G_2$ with probability more than $\frac{2}{3}$. Since $u_f(i_2,j_1)$, $u_f(i_2,j_2)$, $u_f(i_3,j_1)$, and $u_f(i_3,j_2)$ are the same for $G_1$ and $G_2$, the algorithm can only identify $G_1$ and $G_2$ by sampling $u_f(i_1,j_1)$ or $u_f(i_1,j_2)$. Formally, we define the problem as follows. Let $\Omega = [0, 1]^T$ to be the sample space for the outcome of $T$ samples of $r_f(i_1, j_1)$, our goal is to have the following decision rule
\begin{equation}\label{eq:Decision_ruleMain}
    \text{Rule: } \Omega \rightarrow \{ G_1, G_2\},
\end{equation}
\noindent which satisfies the following two properties for any $\omega \in \Omega$:
\begin{equation}\label{eq:DecisionRulePropertyMain}
\begin{split}
    \Pr[u_f \in G_1 | \text{Rule}(\omega) = G_1] > \frac{2}{3} \text{ and }
    \Pr[u_f\in G_2 | \text{Rule}(\omega) = G_2] > \frac{2}{3}.
\end{split}
\end{equation}
\noindent As a result, let $\omega_o \in \Omega$ be the event this $\text{Rule}$ returns $G_1$ (i.e., $\text{Rule}(\omega_o) = G_1$). Then we have:
\begin{equation}\label{eq:DecisionRulePropertyContradictionMain}
    \Pr[u_f \in G_1|\omega_o] - \Pr[u_f \in G_2|\omega_o] > \frac{1}{3}
\end{equation}

Next, for any event $\omega \in \Omega$, let $P_k(\omega) = \Pr[u_f \in G_k|\omega]$ where $k = 1, 2$. Then we have $P_k= P_{k,1} \times \cdots \times P_{k, T}$ where $P_{k, t}$ is the distribution of $t$'th sample of $r_f(i_1, j_1)$. As a result, by applying a basic KL-divergence argument to distributions $P_1$ and $P_2$ \cite[Lemma 2.5]{slivkins2019introduction}, for any event $\omega$ we have $|P_1(\omega) - P_2(\omega)|\leq \epsilon \sqrt{T}$. Plugging $\omega = \omega_0$ and $\epsilon = \frac{1}{3\sqrt{T}}$ we have $|P_1(\omega) - P_2(\omega)|\leq \frac{1}{3}$, contradicting with equation \eqref{eq:DecisionRulePropertyContradictionMain}. 
Therefore, the Stackelberg game instances $G_1$ and $G_2$ in table \ref{table:hardness_learning_new} with $\epsilon = \frac{1}{3\sqrt{T}}$ proves the proposition when $\delta < \Delta$.

\smallskip
\noindent\textbf{$\Omega( 1 )$ lower bound instance when $\delta \geq \Delta$:} 
 
Consider the following two Stackelberg games $G_1$ and $G_2$, where $r_{l}(i,j) \sim \text{Bern}(u_{l}(i,j))$ and $r_{f}(i,j) \sim \text{Bern}(u_{f}(i,j))$, with mean values shown in Table \ref{table:hardness_learning} and $\epsilon \in [0,1]$ is a parameter to be determined. Note that in $G_1$ follower action $j_2$ is dominated by action $j_1$. Thus, we have $\Delta = 0$.

 \begin{table*}
\begin{center}
        \begin{tabular}{|c|  c|  c|}
        \hline
        $u_{l}, u_{f}$ & $j_1$ & $j_2$ \\
        \hline
        $i_1$ & $1, \frac{1+\epsilon}{2} + \delta$ & $0, \frac{1-\epsilon}{2}$\\ 
        \hline
        $i_2$ & $\frac{1}{2}, 1$ & $\frac{1}{2}, 1$\\
        \hline
        \end{tabular}
        \hspace{15mm}
        \begin{tabular}{|c|c|c|}
        \hline
        $u_{l}, u_{f}$ & $j_1$ & $j_2$ \\
        \hline
        $i_1$ & $1, \frac{1-\epsilon}{2} + \delta$ & $0, \frac{1+\epsilon}{2}$\\
        \hline
        $i_2$ & $\frac{1}{2}, 1$ & $\frac{1}{2}, 1$\\
        \hline
        \end{tabular}
\end{center}
\vspace{2mm}
\caption{Example instances $G_1=\{u_l, u_f\}$ (left) and $G_2=\{u_l, u_f\}$ (right). Each table represents a Stackelberg game $u_l, u_f \in \mathbb{R}^{2\times 2}$.}
\label{table:hardness_learning}
\end{table*}

According to Table \ref{table:hardness_learning}, it is straightforward that the $\delta$-RSE of Stackelberg game $G_1$ is  $\langle i_1, j_1 \rangle$ where the leader plays pure strategy $i_1$ and follower responds with pure strategy $j_1$. Leader's utility at $\delta$-RSE is $u_{\textup{RSE}} (\delta) = 1$ for $G_1$. On the other hand, for Stackelberg game $G_2$, we have $u_{f}(\bvec{x}, j_1) - u_{f}(\bvec{x}, j_2) < \delta$ for any leader strategy $\bvec{x}$. Thus, $j_2$ is always in the follower's $\delta$-optimal  response set. To achieve the best utility, the leader plays pure strategy $i_2$ while the follower is indifferent between playing $j_1$ or $j_2$. As a result, $u_{\textup{RSE}}(\delta) = \frac{1}{2}$ for $G_2$.

Suppose the learning algorithm can output leader strategy $\hat{\bvec{x}}$ that violates equation \eqref{eq:observation2}, i.e.,
$$ \min_{j \in \BR_{\delta}(\hat{\bvec{x}})} u_l(\hat{\bvec{x}}, j) > u_{\textup{RSE}}(\delta) - 1/2.$$ 
Then for game $G_1$, we have $\min_{j \in \BR_{\delta}(\hat{\bvec{x}})} u_{l}(\hat{\bvec{x}}, j)>\frac{1}{2}$ while $u_{\textup{RSE}}(\delta) = \max_{\bvec{x} \in \Delta^m}\min_{j \in \BR_{\delta}(\bvec{x})} u_{l}(\bvec{x}, j) =\frac{1}{2}$ for game $G_2$. Then this means the learning algorithm can identify $G_1$ from $G_2$.

Therefore, consider the learning algorithm that samples $r_f(i, j)$ for $T$ times, and the goal is to identify if $u_f \in G_1$ or $u_f \in G_2$. We prove the proposition by contradiction. Suppose Proposition \ref{prop:learningwithgap} is not correct, then with $T$ samples, we can  identify if $u_f \in G_1$ or $u_f \in G_2$ with probability more than $\frac{2}{3}$. Since $u_f(i_2,j_1)$ and $u_f(i_2,j_2)$ are the same for $G_1$ and $G_2$, the algorithm can only identify $G_1$ and $G_2$ by sampling $u_f(i_1,j_1)$ or $u_f(i_1,j_2)$. Formally, we define the problem as follows. Let $\Omega = [0, 1]^T$ to be the sample space for the outcome of $T$ samples of $r_f(i_1, j_1)$ and our goal is to have the following decision rule
\begin{equation}\label{eq:Decision_rule}
    \text{Rule: } \Omega \rightarrow \{ G_1, G_2\}
\end{equation}
which satisfies the following two properties:
\begin{equation}\label{eq:DecisionRuleProperty}
\begin{split}
    \Pr[ u_f \in G_1 | \text{Rule(observations)} = G_1 ] > \frac{2}{3} \\
    \Pr[ u_f\in G_2 | \text{Rule(observations)} = G_2 ] > \frac{2}{3}
\end{split}
\end{equation}
Therefore, let $\omega_o \in \Omega$ be the event this $\text{Rule}$ returns $G_1$. Then we have:
\begin{equation}\label{eq:DecisionRulePropertyContradiction}
    \Pr[u_f \in G_1| \omega_o] - \Pr[u_f \in G_2|\omega_o] > \frac{1}{3}
\end{equation}
Next, for any event $\omega \in \Omega$, let $P_k(\omega) = \Pr[u_f \in G_k|\omega]$ where $k = 1, 2$. Then we have $P_k= P_{k,1} \times \cdots \times P_{k, T}$ where $P_{k, t}$ is the distribution of $t$'th sample of $r_f(i_1, j_1)$. As a result, by applying a basic KL-divergence argument to distributions $P_1$ and $P_2$ \cite[Lemma 2.5]{slivkins2019introduction}, for any event $\omega$ we have $|P_1(\omega) - P_2(\omega)|\leq \epsilon \sqrt{T}$. Plugging $\omega = \omega_0$ and $\epsilon = \frac{1}{3\sqrt{T}}$ we have $|P_1(\omega) - P_2(\omega)|\leq \frac{1}{3}$, contradicting with equation \eqref{eq:DecisionRulePropertyContradiction}. 

Therefore, choosing the problem class to be $G_1$ and $G_2$ in table \ref{table:hardness_learning} with $\epsilon = \frac{1}{3\sqrt{T}}$, finishes the proof.

\qed

\end{document}